\documentclass[runningheads]{llncs}

\usepackage{graphicx}
\usepackage{comment}
\usepackage{xcolor}
\usepackage{makecell}
\usepackage{colortbl}
\usepackage{multirow}
\usepackage[ruled, linesnumbered, commentsnumbered, longend]{algorithm2e}
\SetAlCapNameFnt{\footnotesize}
\SetAlCapFnt{\footnotesize}

\usepackage{silence}
\usepackage{amsmath,amssymb}
\usepackage[hidelinks]{hyperref}

\begin{document}

\title{Empowering Patients for Disease Diagnosis and Clinical Treatment: A Smart Contract-Enabled Informed Consent Strategy}
\titlerunning{Empowering Patients for Disease Diagnosis and Clinical Treatment Process}

\author{
    Md Al Amin \orcidID{0000-0003-1700-7201} \and
        Hemanth Tummala \orcidID{0009-0007-7778-5845}\and
            Rushabh Shah \orcidID{0009-0005-5658-0950} \and
                Indrajit Ray \orcidID{0000-0002-3612-7738}}

\authorrunning{M. Al Amin et al.}

\institute{Computer Science Department, Colorado State University, Fort Collins, CO, USA
    \email{\{Alamin, Hemanth.Tummala, Rushabh.Shah2, Indrajit.Ray\}@colostate.edu}
}

\maketitle             

\begin{abstract}
    Digital healthcare systems have revolutionized medical services, facilitating provider collaboration, enhancing diagnosis, and optimizing and improving treatments. They deliver superior quality, faster, reliable, and cost-effective services. Researchers are addressing pressing health challenges by integrating information technology, computing resources, and digital health records. However, digitizing healthcare introduces significant risks to patient data privacy and security, with the potential for unauthorized access to protected health information. Although patients can authorize data access through consent, there is a pressing need for mechanisms to ensure such given consent is informed and executed properly and timely. Patients deserve transparency and accountability regarding the access to their data: who access it, when, and under what circumstances. Current healthcare systems, often centralized, leave much to be desired in managing these concerns, leading to numerous security incidents. To address these issues, we propose a system based on blockchain and smart contracts for managing informed consent for accessing health records by the treatment team members, incorporating safeguards to verify that consent processes are correctly executed. Blockchain's inherent immutability ensures the integrity of consent. Smart contracts automatically execute agreements, enhancing accountability. They provide a robust framework for protecting patient privacy in the digital age. Experimental evaluations show that the proposed approach can be integrated easily with the existing healthcare systems without incurring financial and technological challenges. 

\keywords{PHI Access\and Diagnosis \and Treatment  \and Treatment Team \and Informed Consent  \and Blockchain \and Smart Contract \and EVM \and Test Network.}
\end{abstract}

\section{Problem Motivations}  \label{sec:problem-motivations} \vspace{-0.8em}
Compared to paper-based systems, electronic health record (EHR) systems make it easier for doctors to work together, make diagnoses more accurate, speed up treatment, and give doctors ready access to patient medical records \cite{menachemi2011benefits}. As healthcare data become more digitized, distributed, and interactive, the healthcare ecosystem is increasingly concerned about the security and privacy of EHR information and systems \cite{keshta2021security}. Several factors contribute to the increased vulnerability of EHR systems. Health workers are often under-trained and under-experienced in handling and securely maintaining information systems.  Software vulnerabilities, security flaws, and human errors allow unauthorized users to access sensitive health records. Insider adversaries can also get their hands on protected medical information, which can cause sensitive patient information to be lost, misused, or shared. Consequently, healthcare providers' responsibility to protect patient privacy and health record confidentiality has increased significantly due to these factors in electronic health data processing \cite{sulmasy2017ethical}.

Advanced security and privacy protocols are needed to prevent unauthorized access and data breach incidents effectively and ensure the security and privacy of patient data. There is ample evidence that shows that improper policy adoption, implementation, and enforcement cause a significant amount of unauthorized access -- without a ``need to know'' -- to EHR data \cite{cda2016regulatory,seh2020healthcare,marchand2020automatic}. Intentionally or unintentionally, access privileges are assigned to users when they should not be. Policies are not followed correctly, and access control rules are not checked or implemented promptly. In some cases, it has been observed that the same roles and privileges are assigned to all employees. Often, individual patient-level policies are not enforced to the word. Additionally, auditing and monitoring practices are deficient, typically initiated only in response to serious complaints or legal obligations. Specifically, these enforcement and specification gaps notably compromise informed consent policies.

Informed consent \cite{parvin2022jurisprudential,lorenzini2022machine} is a critical legal and ethical concept in healthcare. It involves a patient voluntarily agreeing to a medical intervention, procedure, or treatment. This decision comes after the patient is fully informed about the risks, benefits, and alternatives. It also applies to patients agreeing to share their health data for diagnosis, treatment, clinical trials, or research experiments. It enhances patient empowerment, enabling individuals to make well-informed decisions about their health and well-being that align with their best interests. It safeguards patient autonomy and dignity by giving them the authority to direct their health care and treatment choices. Furthermore, it fosters trust between patients and their healthcare providers. The Health Insurance Portability and Accountability Act (HIPAA) in the United States and the General Data Protection Regulation (GDPR) in Europe are just a few of the laws and regulations that regulate informed consent. These legal frameworks are designed to prevent future legal disputes and conflicts by protecting individuals' rights \cite{gopal2023law}. 

Managing informed consent while ensuring the privacy and security of patient health data presents significant challenges. Patients grant permission to a wide range of users, including their treatment team—physicians, nurses, support staff, lab technicians—insurance companies, pharmacists, family members, and other healthcare providers. Patients who require specialized care or consultations with specialists often visit different hospitals. Additionally, life changes such as job relocations or family moves may necessitate transferring to new regions, states, or countries, leading to changes in health plans or insurance coverage. Consequently, patients may need to withdraw previously granted access rights that are no longer necessary or relevant. Emergencies also complicate consent management, as patients might be unable to give immediate permission to access their data. In such cases, designated emergency treatment team members must be able to access the patient's data without prior consent \cite{feinstein2021informed}.

To protect protected health information against unauthorized access and disclosure and to guarantee patients' autonomy over their consent and healthcare resources, it is essential to address the following challenges: individual patient-level policies, known as informed consent, are not adequately captured and enforced. The centralized hospital system presents a single point of failure and acts as the sole source for access audit trails, posing significant risks. There's a significant lack of verifiable consent provenance information, resulting in the absence of an unmodified record of how consent has been executed. Patients need more confidence that their consents are being executed exclusively by designated users, with all other requests being rightfully denied. There is no assurance that consents are executed only under the stipulated conditions; otherwise, access should be denied. Last but not least, patients do not have sufficient control over their consent, impacting their ability to manage access to their health records.

In this paper, we introduce a framework that leverages blockchain technology and smart contracts to enhance the management and enforcement of informed consent \cite{natarajan2017distributed,buterin2014next} to address the above-mentioned challenges and requirements. Our approach is built upon a decentralized distributed ledger technology, specifically using public blockchains like Ethereum \cite{buterin2014next}. This innovative framework ensures automated, secure, and accountable informed consent processes, utilizing smart contracts to automate system operations and safeguard the integrity and accountability of consent. However, it's important to note that our focus is on the mechanism of consent management rather than the security and privacy of the patient's healthcare data. Blockchain technology underpins our system, ensuring that once consent actions are logged, they become immutable, maintaining the audit trail's integrity and detecting unauthorized changes. The inherent security features of blockchain, including nonrepudiation, ensure that participants cannot deny their submissions. Smart contracts play a crucial role by enforcing the execution of informed consent protocols, preventing unauthorized access, and providing real-time notifications through event information for any activities.

To the best of our knowledge, this work is the first to capture and enforce informed consent as the patient-driven policy for disease diagnosis and clinical treatment data access decisions. This paper significantly extends the earlier conference article \cite{al2023informed}. Specifically, this paper makes the following contributions:
\vspace{-.4em}
\begin{itemize} 
     \item [-]Formalizing the sample of the treatment team, protected health information structure, and treatment team member-oriented operations to ensure health data security and privacy by minimizing unnecessary access and disclosures.

    \item [-]Providing mechanisms (Algorithms) for consent administration operations (creation, alteration, termination, withdrawal, and archiving) to keep the patient treatment process uninterrupted and minimize unwanted disclosures.

    \item [-]Proposing graph database-based consent provenance services for users, resources, operations, and conditions for given and executed services to provide consent-related provenance information for transparency and accountability.

    \item [-]Conducting extensive experimental evaluations for the proposed approach on required smart contracts deployment, PPA integrity and informed consent storage and retrieval, and consent administration operations execution.

    \item [-]Performing and analyzing the gas costs, in token and USD, for informed consent and other required smart contracts deployment, storing PPA integrity, and informed consent. Also, analyzing the time requirements for writing data to the blockchain network and reading data from the blockchain network.

    \item [-]Last but not least, as future research directions, delineating a policy compliance framework using informed consent and other applicable components. Using blockchain-based unalterable, accountable, and traceable audit trails as provenance and performing real-time compliance checking employing a blockchain consensus mechanism called Proof of Compliance (PoC).
\end{itemize}
 
The remainder of the paper is organized as follows:  Section \ref{sec:proposed-approach} discusses the proposed framework with the necessary components. Section \ref{sec:consent-enforcement} explains the consent enforcement mechanism. The consent administration operation algorithms are given in Section \ref{sec:consent-administration}.  Section \ref{sec:consent-services} describes consent provenance services. Section \ref{sec:experimental-evaluation} contains the experimental evaluation of the proposed model for deploying smart contracts. Section \ref{sec:additional-factors} includes additional factors to consider to provide services. Some related works are discussed in Section \ref{sec:literature-review}. Section \ref{sec:conclusion-future-directions} concludes the paper with future research directions for healthcare policy compliance framework.

\section{Proposed Framework for Informed Consent} \label{sec:proposed-approach} \vspace{-0.5em}

\begin{figure*}[hbt]
    \centering
    \includegraphics[scale=0.4]{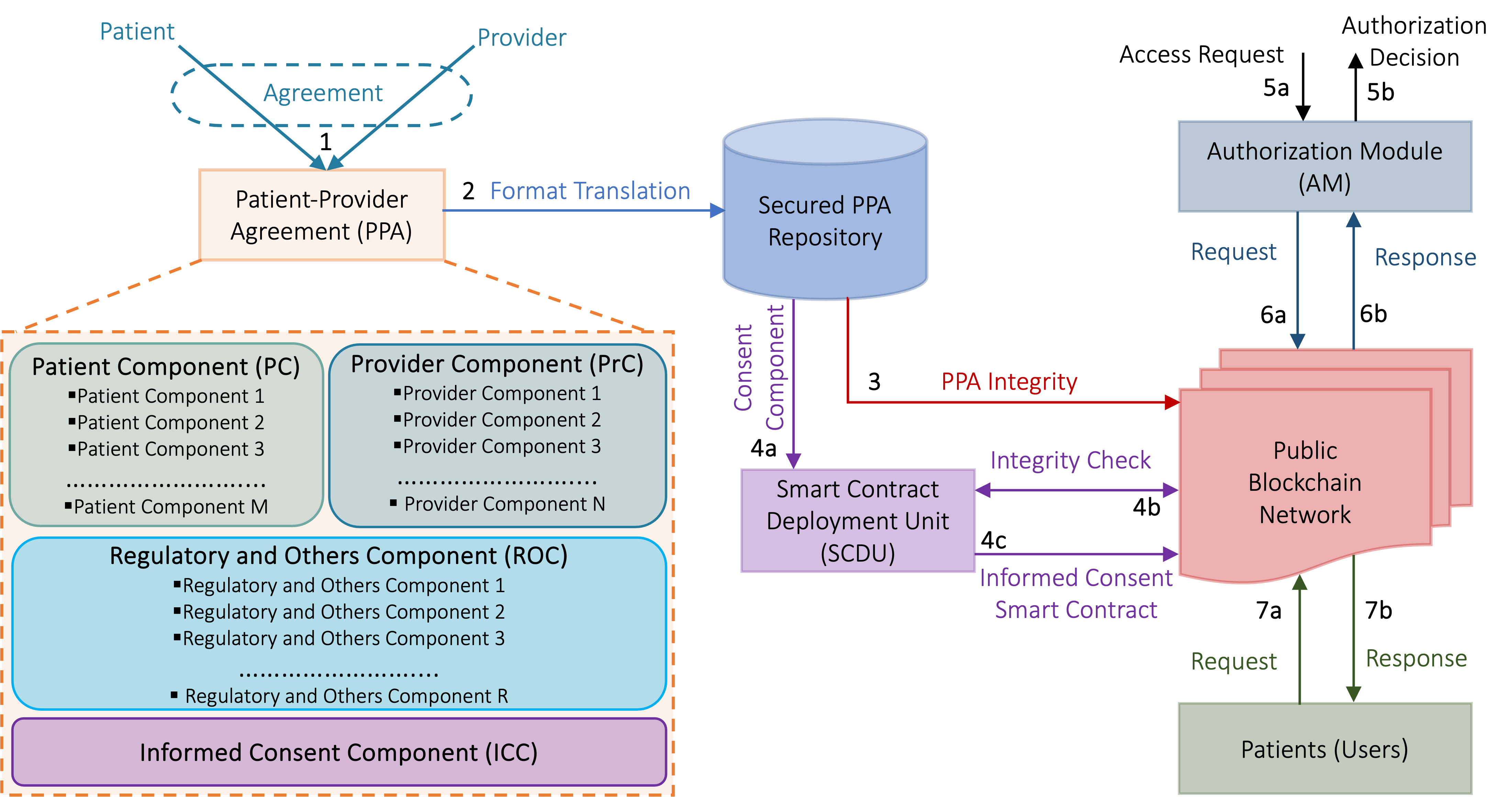}
    \vspace{-2em}
    \caption{Smart Contract-Based Informed Consent Management Framework \cite{al2023informed}.} 
    \label{fig:patient-provider-agreement}
\end{figure*}

The core concept involves incorporating informed consent elements into the patient-provider agreement. Then, create and implement smart contracts for these consents on a public blockchain network. An authorization module triggers these smart contracts upon receiving an access request, detailing the subject, selected operation, targeted objects, and applicable constraints. Upon execution, the smart contract and authorization module's decisions are logged within the blockchain, ensuring a secure and transparent record of consent enforcement. Smart contracts automatically execute and record events and activities as specified, leveraging the blockchain's distributed, immutable, and decentralized nature for secure storage. This ensures that all consents and activities are accurately captured and preserved in accordance with the original patient-provider agreement, maintaining the integrity and unalterability of event logs. Fig. \ref{fig:patient-provider-agreement} shows the smart contract-based informed consent management approach. The subsequent sections explain the patient-provider agreement (Steps \textit{1} to \textit{3}), the informed consent components, and the creation of informed consent smart contracts (Steps \textit{4a, 4b}, and \textit{4c}). Steps \textit{5a, 5b, 6a} and \textit{6b} are discussed in Section \ref{sec:consent-enforcement} and Steps \textit{7a} and \textit{7b} in Section \ref{sec:additional-factors}.

\subsection{Patient-Provider Agreement (PPA)} \vspace{-0.5em}

The patient-provider agreement serves as a framework to clarify the responsibilities and expectations of providing care, aiming to enhance treatment outcomes, minimize risks, and offer better education to patients. These agreements, which can vary significantly from one healthcare organization to another, are tailored to align patient commitments with the specific needs, treatments, and responsibilities dictated by the care and services provided. The healthcare setting, which can range from general hospitals to specialized facilities like emergency rooms, urgent care clinics, dental offices, cancer treatment centers, and physiotherapy clinics, influences the specific elements and structure of a PPA. This customization ensures that the agreement is relevant to the particular nature and requirements of the treatment and services offered. The patient-provider agreement with necessary components is depicted in Fig. \ref{fig:patient-provider-agreement}. A patient-provider agreement formally is composed of four tuples: \[ PPA = (PC, PrC,  ROC, ICC) \] satisfying the following requirements:
\begin{itemize}
    \item[(a)] $PC$ represents a comprehensive collection of patient components, including the patient's personal details, contact data, mailing addresses, pharmacy preferences, billing and insurance particulars, and emergency contacts, among other relevant information. It is the patient's responsibilities to provide and keep this information up-to-date and accurate to get services.
    
    \item[(b)]  $PrC$ encompasses a defined group of provider components, which cover aspects like the treatment team creating, anonymized data sharing, prescription handling, and more. The treatment team consists of medical professionals such as doctors, nurses, support staff, lab technicians, and billing officers, all of whom are integral to managing the patient's care. This set ensures that all aspects of patient treatment process, including medical procedures, required services, insurance coverage, and billing processes, are comprehensively addressed throughout the treatment duration.
 
    \item[(c)] $ROC$ represents a specific collection of regulatory and miscellaneous components, incorporating security and privacy policies that adhere to compliance requirements set by local and federal governments, as well as regulatory bodies like HIPAA and GDPR. It ensures alignment with national and international regulations to protect patient data and uphold privacy standards.

    \item[(d)] $ICC$ is a finite set of informed consent components. It indicates the permission given by the patient to access healthcare data. This work mainly focuses on $ICC$, including (i) identifying, capturing, and storing consent components, (ii) enforcing consents while making access authorizations, (iii) providing mechanisms for consent administration like creation, alteration, termination, expiration, and archiving, (iv) providing consent provenance services for both given and executed consents for users, resources, operations, and conditions. It does not consider and discuss $PC$, $PrC$, and $ROC$. 
\end{itemize}
  
 A PPA is created when a patient visits a hospital, with its validity constrained by a predefined duration. Algorithm \ref{alg:patient-provider-agreement} shows the step-by-step instructions for creating a PPA with $PC$, $PrC$, $ROC$, and $ICC$. A single patient might be associated with multiple PPAs to facilitate the provision of healthcare services. When access requests involve such contracts, the authorization module evaluates them with other applicable policies and components to make decisions. As depicted in Fig. \ref{fig:patient-provider-agreement}, the proposed framework stores the integrity of a PPA on the blockchain, enabling the detection of any deliberate or accidental modifications.

\RestyleAlgo{ruled}
\SetKwComment{Comment}{/* }{ */}
\begin{algorithm}[tb]
    \scriptsize
    \SetKwInOut{KwData}{Input}
    \SetKwInOut{KwResult}{Result}
    \DontPrintSemicolon
\caption{Patient-Provider Agreement (PPA) Creation \cite{al2023informed}}\label{alg:patient-provider-agreement}

\KwData{(i) $PC$, (ii) $PrC$, (iii) $ROC$, (iv) $ICC$, (v) $\mathbb{R}_{PPA}$, (vi) $\mathbb{BN}_{SC}$}
                \textcolor{blue}{\Comment*[r]{$\mathbb{R}_{PPA}$: secured PPA repository, $BN_{SC}$: blockchain network smart contract}}
\KwResult{A formal $PPA$}

\textbf{Input Parameters Initialization} \vfill 
  $PPA_i \gets  \{ PC_i, PrC_i, ROC_i, ICC_i\}$ for patient identity $i$ \vfill
    \quad \textit{(i)} $PC \gets \{ \wp_1, \wp_2, \wp_3,\wp_4, \wp_5,........\wp_M \}$\; \vfill
    \quad \textit{(ii)} $PrC \gets \{ \delta_1, \delta_2, \delta_3,\delta_4, \delta_5,........\delta_N \}$\; \vfill
    \quad \textit{(iii)} $ROC \gets \{ \Re_1, \Re_2, \Re_3,\Re_4, \Re_5,........\Re_P \}$\; \vfill
     \quad \textit{(iv)} $ICC \gets \{ 	\ell_1, \ell_2, \ell_3,\ell_4, \ell_5,........\ell_R \}$\; \vfill
 
 \textbf{PPA Components Integrity Calculation} \vfill        \textcolor{blue}{\Comment*[r]{$\mathbb{H}(\partial)$ calculates hash of $\partial$}}
    \quad \textit{(a)} $\mathbb{H}_{PC} \gets \{ \wp_1, \wp_2, \wp_3,\wp_4, \wp_5,........\wp_M \}$\; \vfill
    \quad \textit{(b)} $\mathbb{H}_{PrC} \gets \{ \delta_1, \delta_2, \delta_3,\delta_4, \delta_5,........\delta_N \}$\;  \vfill
    \quad \textit{(c)} $\mathbb{H}_{ROC} \gets \{ \Re_1, \Re_2, \Re_3,\Re_4, \Re_5,........\Re_P \}$\;\vfill
    \quad \textit{(d)} $\mathbb{H}_{ICC} \gets \{ 	\ell_1, \ell_2, \ell_3,\ell_4, \ell_5,........\ell_R \}$\;  \vfill
    \quad \textit{(e)} $\mathbb{H}_{PPA_i} \gets  \mathbb{H}( \mathbb{H}_{PC}, \mathbb{H}_{PrC}, \mathbb{H}_{ROC}, \mathbb{H}_{ICC})$\; \vfill

 \textbf{PPA Finalization} \vfill 
  \eIf{$PPA_i$ is complete}{
        \textcolor{blue}{\Comment*[r]{complete: presence of $PC$, $PrC$, $ROC$, $ICC$}}

            \eIf{$(\mathbb{R}_{PPA} + PPA_i)$ contains no conflicts}{
                   \textit{(i)} do $\mathbb{R}_{PPA} \gets (\mathbb{R}_{PPA} + PPA_i)$\;
                   \textit{(ii)} add $\mathbb{ID}_{PPA_i}$ to patient profile, $\mathbb{P}_i$\;
                   \textit{(iii)} call $\mathbb{BN}_{SC} (\mathbb{ID}_{PPA_i}, \mathbb{H}_{PPA_i})$\; 
                        \textcolor{blue}{ \Comment*[r]{later PPA integrity verification}}
                   
                   \textbf{\textit{Return: }} Success ($PPA_i$ added to $\mathbb{R}_{PPA}$)\;
            }{
                \textbf{\textit{Error: }} $(\mathbb{R}_{PPA} + PPA_i)$ contains conflicts\;
                            \textcolor{blue}{ \Comment*[r]{ $PPA_i$ revision required to add to $\mathbb{R}_{PPA}$}}
            }
   }{
        \textbf{\textit{Error: }}$PPA_i$ cannot be created\;   \textcolor{blue}{\Comment*[r]{incomplete patient-provider agreement (PPA)}}
 }
\end{algorithm}

\subsection{Informed Consent Structure}

\begin{figure*}[tb]
    \centering
    \includegraphics[scale=0.4]{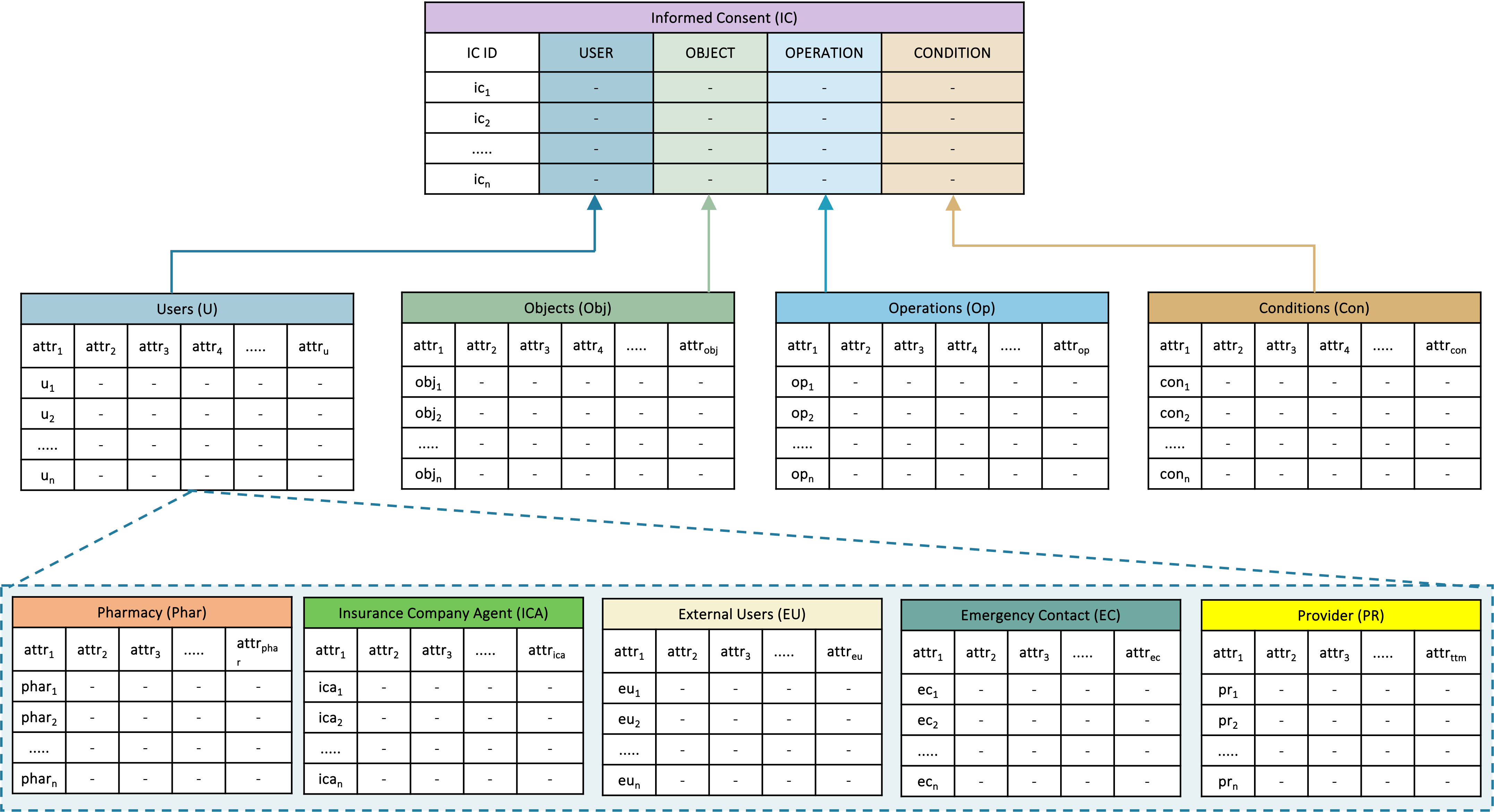}
    \vspace{-2em}
    \caption{Informed Consent Components \cite{al2023informed}.} \label{fig:informed-consent-components}
\end{figure*}

Before granting consent, patients must be fully informed and understand all aspects of their specific consent. Fig. \ref{fig:informed-consent-components} shows the informed consent conceptual structure. The informed consent formally is composed of four tuples: \[ IC = (U, O,  OP, CON) \]  satisfying the following requirements:
\begin{itemize}
    \item[(1)] $U$ represents a finite set of authorized users, expressed as $\lbrace u_1, u_2, u_3, .....\rbrace$. Users within this set are permitted to execute specific operations on healthcare resources, contingent upon fulfilling predefined conditions.
    
    \item[(2)]  $O$ signifies a finite set of protected healthcare information, referred to as protected objects. This set is denoted by $\lbrace o_1, o_2, o_3, .....\rbrace$ encapsulating the various items of sensitive health data safeguarded under privacy regulations.
    
    \item[(3)] $OP$ is a finite set of permissible actions, represented as $\lbrace op_1, op_2, op_3, ...\rbrace$. These operations define the range of system activities that authorized users may perform on protected health information, with common examples including reading, writing, and updating data.

    \item[(4)] $CON$ represents a finite collection of prerequisites, outlining the specific conditions that must be met by users to engage with protected health information. This set of conditions is formally denoted as $\lbrace con_1, con_2, con_3, ...\rbrace$, serving as the criteria governing user actions on sensitive healthcare data.
\end{itemize}

The healthcare system encompasses diverse users, each with distinct roles and responsibilities in executing their duties. A patient's treatment team may consist of doctors, nurses, support staff, laboratory technicians, billing officers, the patient's designated emergency contact, and other hospital personnel appointed by the supervising authority. All aspects, including medical care, insurance coverage, and billing processes, are accounted for throughout a patient's treatment duration. Informed consent users can be anyone from five groups of people: \textit{(i) provider, (ii) emergency contract, (iii) external users, (iv) insurance company agent,} and \textit{(v) pharmacy}. Table \ref{table:patient-treatment-team-member} shows a sample structure of treatment team members and their responsibilities regarding the treatment process. External users are from different hospitals when a patient is transferred for better treatment if the situation demands it. Usually, they have temporary access to admitted patients' health records. For this study, we don't consider external users as treatment team members, as shown in Table \ref{table:patient-treatment-team-member}.

\begin{table}[tb!]
    \caption{Patient Treatment Team Members and Responsibilities.} \label{table:patient-treatment-team-member}
    \vspace{-.8em}
    \resizebox{\columnwidth}{!}{
        \begin{tabular}{|c|l|l|} 
            \hline
            \rowcolor{lightgray} \textbf{SN} & \textbf{Treatment Team Member}    &  \textbf{Responsibilities} \\
            \hline
                1 & Doctor (DOC) & Viewing patient's healthcare data\\
            \hline 
                2 & Nurse (NRS) & Creating new patient's healthcare data\\
            \hline 
                3 & Support Staff (STF) & Correcting erroneous or appending patient's healthcare data\\
            \hline 
                4 & Billing Officer (BLO) & Viewing patient's healthcare data\\
            \hline 
                5 & Radiology Lab Tech (RLT) & Creating new patient's healthcare data\\
            \hline 
                6 & Pathology Lab Tech (PLT) & Correcting erroneous or appending patient's healthcare data\\
            \hline 
                7 & Emergency Contact (EMC) & Viewing patient's healthcare data\\
            \hline 
                8 & Pharmacist (PHR) & Creating new patient's healthcare data\\
            \hline 
                9 & Insurance Agent (INA) & Correcting erroneous or appending patient's healthcare data\\
            \hline 
        \end{tabular}
    }
\end{table}

The digital record of a patient's medical history that the healthcare provider maintains is what the term "object" refers to in healthcare. This record encapsulates comprehensive administrative and clinical data relevant to the patient's care, including demographics, progress notes, diagnoses, prescriptions, vital statistics, past medical history, vaccinations, lab results, and imaging reports. The imperative to safeguard these records from unauthorized access underscores the critical role of informed consent, which empowers patients to grant specific permissions to users for performing designated operations on their protected health information. Table \ref{table:protected-health-inforrmation} shows the health records, categorized by ID, name, and description, illustrating the scope of data managed within the system.

\begin{table*}[tb]
\centering
\caption{Sample Patient Protected Health Information (PHI) Structure.} \label{table:protected-health-inforrmation}
\vspace{-.8em}
\scriptsize
\begin{tabular}{|c| l| l|} 
\hline
\rowcolor{lightgray}  \textbf{PHI ID} &  \textbf{PHI Name} & \textbf{PHI Description}\\
\hline
 PHI1001 & Demographic Info & Patient's information\\ 
 \hline 
 PHI1002 &  Previous Medical History & Old medical records from another hospital\\
\hline
 PHI1003 & Immunizations  & Immunization records that are administered over time\\
\hline
 PHI1004& Allergies & Various allergies sources, triggering condition, remediation\\
\hline
 PHI1005 & Visit Notes & Physiological data, disease description, advice, follow-up\\
\hline
 PHI1006& Medications \& Prescription & Prescribed medications including name, dosage, etc.\\
\hline
 PHI1007 & Pathology Lab Works & Blood work\\
\hline
 PHI1008 & Radiology Lab Works & Imaging and Radiology Lab results\\
\hline
 PHI1009 & Billing and Insurance & Bank account and insurance policy Information\\
 \hline
 PHI1010 & Payer Transactions & Bills of doctor visit, lab works, and medications\\
\hline
\end{tabular}
\end{table*}

In the healthcare sector, authorized personnel perform various operations to provide treatment and services. This research focuses on three primary operations: \textit{reading}, \textit{writing}, and \textit{updating}. The \textit{reading} operation allows users to access healthcare records, provided their requests are valid and follow all relevant policies, maintaining data integrity without altering the data's state. Yet, this could breach confidentiality and privacy if access is granted inappropriately. \textit{Writing} operations, in contrast, result in the creation of new data within the records. \textit{Updating} operations, conversely, are utilized to modify existing data, whether to append new information or correct errors, without generating new data. Writing and updating operations must maintain data integrity, necessitating rigorous policy enforcement. To illustrate the distribution of access rights, a sample of protected health information for patients is detailed in Table \ref{table:protected-health-inforrmation}. In contrast, Table \ref{table:patient-treatment-team-member} lists the members of a patient's treatment team. Furthermore, Table \ref{table:user-oriented-health-records-operations} delineates the permissible PHI operations for each treatment team member, indicating that not all members have access to all forms of PHI for performing their job responsibilities. In addition to the treatment team, the patient also has the rights to read, write, and update specific health records. This measure is essential for maintaining security and privacy. However, this study doesn't provide the informed consent mechanism for patient access. We assume that patient access is adequately controlled.

\begin{table*}[tb]
\centering
\caption{Treatment Team Member Oriented PHI Operations.} \label{table:user-oriented-health-records-operations}
\vspace{-0.8em}
\scriptsize
\begin{tabular}{|c |  l |  l | l |} 
\hline
\rowcolor{lightgray} \textbf{PHI ID} & \textbf{Read Operation} & \textbf{Write Operation} & \textbf{Update Operation}    \\
\hline
 PHI1001 &  \textit{Patient}, DOC, STF, EMC &  \textit{Patient}, STF &  \textit{Patient}, STF \\
 \hline 
 PHI1002 & DOC, \textit{Patient} & \textit{Patient}, DOC & \textit{Patient}, DOC \\
\hline
 PHI1003 & DOC, \textit{Patient}, PLT & PLT & PLT\\
\hline
 PHI1004 &  DOC, \textit{Patient}, NRS & \textit{Patient}, PLT & \textit{Patient}, PLT\\
\hline
 PHI1005 & DOC, NRS, \textit{Patient}, EMC & DOC & DOC\\
\hline
 PHI1006&  DOC, \textit{Patient}, NRS, PHR, INA, EMC & DOC & DOC\\
\hline
 PHI1007 &  PLT, DOC, \textit{Patient}, EMC & PLT & PLT\\
\hline
 PHI1008 &  RLT, DOC, \textit{Patient}, EMC &  RLT & RLT\\
\hline
 PHI1009 & \textit{Patient}, BLO, INA & BLO, \textit{Patient} & BLO, \textit{Patient}\\
 \hline
 PHI1010 &  \textit{Patient}, BLO, INA & BLO, INA  &  BLO, INA \\
\hline
\end{tabular}
\end{table*}

Informed consent enforcement, rejection, or revocation may be subject to various requirements or conditions to provide treatment and services. These conditions, while not exhaustive, may include, but are not limited to:
\vspace{-.6em}
\begin{itemize}
    \item[(i)] \textbf{\textit{Time Limitations:}} These limitations dictate that access to a patient's healthcare data is permissible only during predefined intervals. Consider a scenario where consent is conditional upon standard office hours, from 8 AM to 5 PM. Access requests made outside these hours are automatically denied. Any such attempt to access the patient's records is documented as part of the system's audit trail for record-keeping and review purposes.
    
    \item[(ii)] \textbf{\textit{Calendar Restrictions:}} These restrictions confine access to healthcare data to specific dates. For instance, if consent is granted until the 30th of June, any access request submitted on or after the 1st of July will be denied. This ensures that patient-protected health information is only accessible before the consent's expiration date.

    \item[(iii)] \textbf{\textit{Daily Access Parameters:}} Access to patient data can be regulated based on specific days, such as weekdays (Monday through Friday), weekends (Saturday and Sunday), or public holidays. This means that authorization to access data is aligned with the assigned days. For example, a primary care physician may be granted access to patient records during their working days, from Monday to Friday. Conversely, access would be restricted on weekends, reflecting the physician's off-duty status.

    \item[(iv)] \textbf{\textit{Geographical Access Restrictions:}} These restrictions are predicated on the user's location, permitting data access only within predefined geographical boundaries. For instance, healthcare practitioners may access patient information exclusively within the confines of a hospital, or more specifically, only when situated in critical care areas such as the emergency room, ensuring immediate and location-relevant use of data for patient treatment.

    \item[(v)] \textbf{\textit{IP-based Constraints:}} This stipulation restricts access to healthcare resources to devices with pre-approved IP addresses or known and trusted networks. Access attempts from IP addresses that are not on the designated allowlist are automatically denied, ensuring that only verified and recognized devices within the healthcare network can access patient data.

    \item[(vi)] \textbf{\textit{Access Frequency Limits:}} It limits the number of times a user is permitted to access particular health record. For example, an external physician may be granted consent to review a patient's records a maximum of five times. After the physician has accessed five times, the consent automatically expires, barring further access until renewed consent is obtained or given.
\end{itemize}

\vspace{-.3em}
The conditions specified herein are not exhaustive and have been selected based on their relevance to this study. It is acknowledged that additional conditions may arise contingent upon various factors, such as the specific treatment being administered, patient demographics, the healthcare provider's operational policies, and other situational requirements. In an era where advanced technology can be a double-edged sword, it is plausible for malicious entities to manipulate or fabricate conditions to access healthcare data or compromise credentials illicitly. To counteract such threats, it is imperative to implement robust, multi-layered security protocols. These protocols must be rigorously designed to verify the authenticity and integrity of condition credentials, ensuring they remain impervious to unauthorized alteration or deception.

\subsection{Informed Consent Smart Contract Generation} \vspace{-.5em}

Once a patient-provider agreement or PPA is created and stored in the repository, all informed consent components are deployed as smart contracts. Steps \textit{4a}, \textit{4b}, and \textit{4c} in Fig. \ref{fig:patient-provider-agreement} show the process. The authorization module needs to access these smart contracts, integrating them into the decision-making process alongside other required components. These components include subject, object, operational attributes, environmental conditions, as well as organizational, regulatory, and additional policies as necessary. In this approach, a single smart contract  is maintained that acts as a consent container. If there is no contract then a new is created and added to the patient profile and hospital systems for services. The contract address is an identifier for a smart contract in the blockchain to store and retrieve informed consents. This smart contract comprises both functions and data, structured into two distinct data units: the consent repository and the consent archive (Fig. \ref{fig:patient-smart-contract-structure}). The repository holds active informed consents, accessible to the authorization unit for processing access requests. Conversely, the archive stores inactive, read-only historical consents. They are not executable for current authorizations and are crucial for compliance verification and resolving disputes in investigative or legal contexts.

The smart contract deployment unit, \textit{SCDU}, collects all consent components from PPA and checks integrity to confirm that collected consents are not modified deliberately or inadvertently. In step 3 in Fig. \ref{fig:patient-provider-agreement}, PPA integrity as the hash from Algorithm \ref{alg:patient-provider-agreement} ($\mathbb{H}_{PPA_i}$) is stored in the blockchain network along with \textit{PPA ID}. To verify PPA integrity, \textit{SCDU} calls the corresponding smart contract function to retrieve the PPA integrity value stored in the network. After receiving, it compares with the current integrity from the PPA repository. Any modification of consent components voids the consent. If there is no modification, then SCDU creates and deploys smart contract(s) to the blockchain network.  The SCDU works as a secure and trusted API designed to maintain the integrity of consent components without modification. It also ensures that no consent-related information is disclosed to any unauthorized entities. However, this paper doesn't provide detailed architecture and functional mechanisms of SCDU.

\section{Informed Consent Enforcement for PHI Authorization} \label{sec:consent-enforcement} \vspace{-1em}

\begin{figure*}[hbt]
    \centering
    \includegraphics[width=\linewidth]{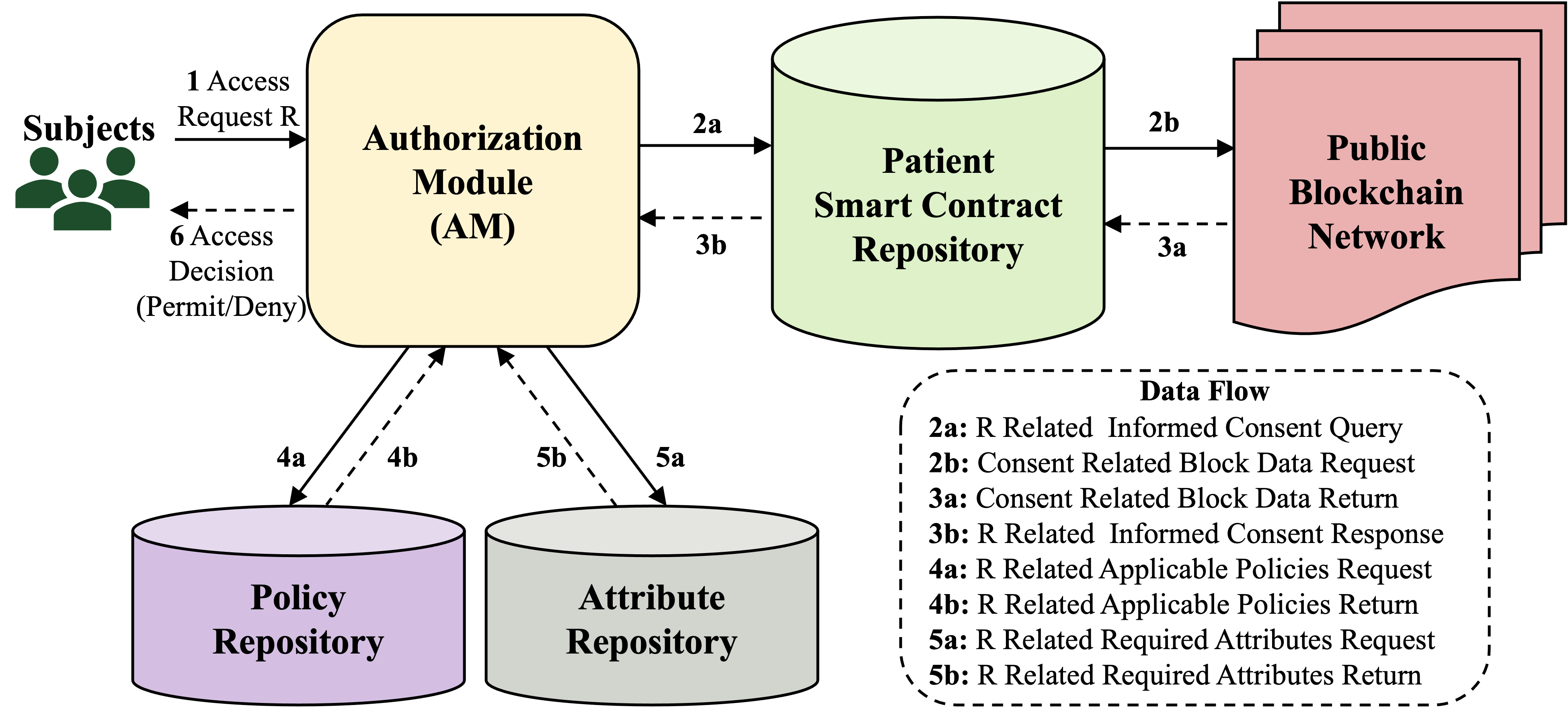}
    \vspace{-2em}
    \caption{Informed Consent Enforcement Process for PHI Access Authorization.} \label{fig:consent-enforcement}
\end{figure*}

Capturing and storing informed consent is not enough. There must be some mechanisms for enforcing consent for making PHI access authorization decisions for the treatment team member. Consent enforcement ensures that related consents are executed while making access decisions for the requests. In the proposed model, all consents are stored on the public blockchain network as smart contracts and cannot be enforced until they are called. The authorization module (AM) considers all applicable consents from a patient while making an authorization decision for access requests. The AM also considers applicable policies and required attributes. The attributes can be subject, object, operation, and environmental attributes. Fig. \ref{fig:consent-enforcement} shows the consent enforcement process.

When a subject sends an access request (R) in Step \textit{1}, the Authorization Module (AM) checks the blockchain by contacting the patient's smart contract (Steps \textit{2a} and \textit{2b}) to find information on the informed consent related to the request (Steps \textit{3a} and \textit{3b}). It also looks up policies related to the request in Steps \textit{4a} and \textit{4b} and gathers the necessary attributes in Steps \textit{5a} and \textit{5b}. After reviewing the consent, policies, and attributes, the AM decides whether to grant access. This decision is returned to the subject in Step \textit{6}. Following the decision, the AM records details of the decision-making process on the blockchain as event logs through the smart contract. 

This study assumes the authorization module is secure and has not been tampered with. The communication between the AM and the blockchain is protected against attacks by malicious users. We don't provide the detailed structure of the authorization module. However, we refer the interested readers seeking a deeper understanding of the module's functionalities to the Attribute-Based Access Control (ABAC) model, as detailed in the work of Hu et al. \cite{hu2013guide}.

\section{Consent Administration} \label{sec:consent-administration} \vspace{-1em}
This section briefly explains the various operations involved in consent administration, such as consent creation, alteration, termination, expiration, and archiving. Ensuring these operations are carried out without introducing privilege conflicts, leakages, or incomplete treatment teams is crucial. The most important aspect of these operations is ensuring they do not disrupt the treatment process. For example, if consent from a pharmacy agent is withdrawn, the agent cannot access or process the patient's prescription to provide medications, which can cause delays in treatment and ultimately lead to life-threatening consequences. The consent owner or patient must invoke consent modification and termination functions. Additionally, consent expiration and archiving operations should be performed automatically as default functions when the conditions are met. Consent management is a complex, multi-step process that must be carefully thought out to ensure efficiency and meet all relevant standards and requirements.

\textbf{Consent Creation: } This process involves generating new consent, with complete details and functionalities outlined in Section \ref{sec:proposed-approach}, including the necessary components and their interplay. New consents can be formulated either during or after the Patient-Provider Agreement (PPA) is established to accommodate the addition of new treatment team members. However, integrating new consents may lead to conflicts with existing ones, such as an incomplete treatment team. Therefore, a thorough check is required to avoid conflicts or issues before adding consent to the patient smart contract. After successful verification, the consent is deployed as a smart contract. The procedure is encapsulated in Algorithm \ref{alg:consent-creation}, detailing the sequential actions required for successful consent creation.

\begin{algorithm}[htb]
    \caption{Consent Creation} \label{alg:consent-creation}
    \scriptsize
    \SetKwInput{KwInput}{Input}
    \SetKwInput{KwOutput}{Output}
    \DontPrintSemicolon
    
  \KwInput{(i) $\mathbb{IC}_{New}:$ New Informed Consent, (ii) $\mathbb{R}_{IC}:$ Informed Consent Repository}
  \KwOutput{Success or failure status}
    \textbf{Consent Creation} \vfill 
        \eIf{($\mathbb{R}_{IC}$ + $\mathbb{IC}_{New}$) contains no conflicts}{
                \textcolor{blue}{\tcc{conflicts mean incomplete treatment team or process, leakage/contradictions}}
                
                \eIf{$\mathbb{R}_{IC} \xleftarrow{} (\mathbb{R}_{IC}+ \mathbb{IC}_{New})$ == True}{
                    \KwRet \textit{success: $\mathbb{IC}_{New}$ is added to $\mathbb{R}_{IC}$}\;
                            \textcolor{blue}{\tcc{$\mathbb{IC}_{New}$ is ready to be executed for authorizations}}
                    }{
                         \KwRet \textit{error: $\mathbb{IC}_{New}$ is not added to $\mathbb{R}_{IC}$}\; 
                                \textcolor{blue}{\tcc{$\mathbb{IC}_{New}$ must be modified and tested to be added to $\mathbb{R}_{IC}$}}
                    }
             }{
            \KwRet \textit{error: modify/update $\mathbb{IC}_{New}$}\;
                    \textcolor{blue}{\tcc{avoid leakage/contradictions}}
        }
\end{algorithm}

\textbf{Consent Alteration: } There are times when it's necessary to update a consent for various reasons, such as correcting errors, modifying current users, objects, or conditions, adding new users, entities, or conditions, dropping users, objects, or conditions, and others. The old consent is added to the consent archive if any modification occurs. The complete process is described in Algorithm \ref{alg:consent-alteration} with all the necessary components and operations. When giving or taking consent, there's a chance of unintentionally creating errors, which could lead to unwanted events, including security incidents. Once a mistake is realized, it's crucial to resolve it immediately to avoid any undesirable incidents or PHI disclosures.

\begin{algorithm}[tb]
    \SetKwInput{KwInput}{Input}
    \SetKwInput{KwOutput}{Output}
    \DontPrintSemicolon
    \scriptsize
    \caption{Consent Alteration}\label{alg:consent-alteration}
    
   \KwInput{(i) $\mathbb{IC}_{Old}:$ Old Informed Consent ID, (ii) $\mathbb{IC}_{New}:$ New Informed Consent, (iii) $\mathbb{R}_{IC}:$ Informed Consent Repository, (iv) $\mathbb{AR}_{IC}:$ Informed Consent Archive}
  \KwOutput{Success or failure status}

 \textbf{Consent Modification} \vfill         
        \eIf{($\mathbb{R}_{IC}$ - $\mathbb{IC}_{Old}:$) contains no conflicts}{
                \textcolor{blue}{\tcc{leakage/contradictions}}
            \eIf{no conflict is in ($\mathbb{R}_{IC}$ - $\mathbb{IC}_{Old}$ + $\mathbb{IC}_{New}$)}{
                \eIf{$\mathbb{R}_{IC} \xleftarrow{} (\mathbb{R}_{IC}+ \mathbb{IC}_{New})$ == True}{
                    \textit{(i)} do $\mathbb{AR}_{IC} \gets (\mathbb{AR}_{IC} + {IC}_{Old})$\;
                    \textit{(ii)} add $\mathbb{IC}_{New}$ to patient profile\;
                    \KwRet \textit{success: $\mathbb{IC}_{New}$ added to $\mathbb{R}_{IC}$}\;
                            \textcolor{blue}{\tcc{$\mathbb{IC}_{Old}$ cannot be executed}}
                            \textcolor{blue}{\tcc{$\mathbb{IC}_{New}$ can be executed now}}
                    }{
                         \KwRet \textit{error: $\mathbb{IC}_{New}$ is not added to $\mathbb{R}_{IC}$}\; 
                               
                    }
            } {
                    \KwRet \textit{error: modify/update $\mathbb{IC}_{New}$}\;
                    \textcolor{blue}{\tcc{avoid leakage/contradictions}}
            }
                
             }{
            \KwRet \textit{error: $\mathbb{IC}_{Old}:$ cannot be modified}\;
                    \textcolor{blue}{\tcc{avoid leakage/contradictions}}
        }         
\end{algorithm}

\textbf{Consent Termination: } Consent withdrawal occurs when patients decide to halt their data sharing. Additionally, if consent is erroneously assigned or contains onerous conditions, it can be rescinded by the patient or the overseeing hospital authority. Upon revoking consent, it is imperative to inform all relevant parties. The revoked consent is then documented in an archive, addressing any subsequent legal or regulatory inquiries. Algorithm \ref{alg:consent-termination} shows the step-by-step instructions for termination operation. It's important to note that if the revoked consent is critical to ongoing treatment, its removal could lead to severe repercussions, such as disruptions in care or medication availability or services.

\begin{algorithm}[tb]
    \SetKwInput{KwInput}{Input}
    \SetKwInput{KwOutput}{Output}
    \DontPrintSemicolon
    \scriptsize
    \caption{Consent Termination} \label{alg:consent-termination}
  \KwInput{(i) $\mathbb{ID}_{IC}$: Informed Consent ID, (ii) $\mathbb{R}_{IC}:$ Informed Consent Repository, (iii) $\mathbb{AR}_{IC}:$ Informed Consent Archive}
  \KwOutput{Success or error status}
        \eIf{$\mathbb{ID}_{IC}$ is in $\mathbb{R}_{IC}$}
           {    
                \eIf{ no conflict is in $\mathbb{R}_{IC}-\mathbb{ID}_{IC}$} {
                          \textit{(i)} do   $\mathbb{R}_{IC} \xleftarrow{} (\mathbb{R}_{IC} - \mathbb{ID}_{IC})$\;
                                    \textcolor{blue}{\tcc{delete selected informed consent from repository}}
                           \textit{(ii)} do  $\mathbb{AR}_{IC} \xleftarrow{} (\mathbb{AR}_{IC} + \mathbb{ID}_{IC})$\;
                                    \textcolor{blue}{\tcc{add deleted informed consent to archive}}
                            \KwRet \textit{success: } $\mathbb{ID}_{IC}$ is terminated from $\mathbb{R}_{IC}$\;
                                        \textcolor{blue}{\tcc{$\mathbb{IC}_{Old}$ cannot be executed}}
                } {
                    \KwRet \textit{error}\;
                                \textcolor{blue}{\tcc{$\mathbb{R}_{IC}-\mathbb{ID}_{IC}$ contains conflict}}
                }
            }{
              \KwRet \textit{error}\; 
                    \textcolor{blue}{\tcc{$\mathbb{ID}_{IC}$ does not exist in $\mathbb{R}_{IC}$ }}
            }           
\end{algorithm}

\textbf{Consent Expiration: } Consent may be invalidated if predefined conditions are not met, such as specific dates or access limits. For instance, if a doctor is granted consent to access a patient's data up to five times, this consent automatically expires upon the fifth access. Any attempt to access the data a sixth time would be unauthorized due to the expiration of consent. Consent conditions, including access frequency and others, are designed to maintain its validity. Algorithm \ref{alg:consent-expiration} presents a set of instructions for the expiration process from initiation to completion. The system must monitor these conditions automatically to prevent delays and oversights, ensuring efficient and accurate consent management.

\begin{algorithm}[tb]
    \SetKwInput{KwInput}{Input}
    \SetKwInput{KwOutput}{Output}
    \SetKw{KwBy}{by}
    \DontPrintSemicolon
    \scriptsize
    \caption{Consent Expiration}  \label{alg:consent-expiration}
  \KwInput{(i) $\mathbb{CON}_{IC}$: Informed Consent Conditions, (ii) $\mathbb{R}_{IC}$: Informed Consent Repository, (iii) $\mathbb{AR}_{IC}$: Informed Consent Archive}
  \KwOutput{Success or error status}
  \textbf{Consent Expiration} \vfill 
        \For{$con \gets \mathbb{CON}_{IC_{Start}}$ \KwTo $\mathbb{CON}_{IC_{End}}$ \KwBy $1$}{
                \For{$ic \gets \mathbb{R}_{IC_{Start}}$ \KwTo $\mathbb{R}_{IC_{End}}$ \KwBy $1$}{
                            \eIf{$con$ is not satisfied by $ic$}
                                    {
                                      \textit{(i)} do  $\mathbb{R}_{IC} \xleftarrow{} (\mathbb{R}_{IC} - ic)$\;
                                                \textcolor{blue}{\tcc{delete expired informed consent from repository}}
                                      \textit{(ii)} do $\mathbb{AR}_{IC} \xleftarrow{} (\mathbb{AR}_{IC} + ic)$\;
                                                \textcolor{blue}{\tcc{add expired informed consent to archive}}
                                    }{
                                    
                                    }
                }{
                
                }
        }{
        
        }          
\end{algorithm}

\textbf{Consent Archiving: } This procedure moves modified, withdrawn, and expired consents into a read-only archive or repository. It ensures no consent within this database remains active and cannot be enforced for protected health information access authorization. The archive's primary objective is maintaining a record of consents for addressing legal or regulatory queries and facilitating policy compliance verification, considering that certain operations might have been conducted under these consents. Furthermore, it allows patients to view all their historical consents, including any altered, revoked, or expired. The consent archiving process is delineated into steps in Algorithm \ref{alg:consent-archiving}.

\begin{algorithm}[tb]
    \SetKwInput{KwInput}{Input}
    \SetKwInput{KwOutput}{Output}
    \SetKw{KwBy}{by}
    \DontPrintSemicolon
    \scriptsize
    \caption{Consent Archiving for Alteration, Termination, and Expiration}  \label{alg:consent-archiving}
  \KwInput{(i) $\mathbb{ID}_{IC}$: Informed Consent ID, (ii) $\mathbb{R}_{IC}$: Informed Consent Repository, (iii) $\mathbb{AR}_{IC}$: Informed Consent Archive}
  \KwOutput{Success or error status}
         \textbf{Consent Archiving} \vfill         
        \eIf{($\mathbb{R}_{IC}$ - $\mathbb{ID}_{IC}:$) contains no conflicts}{
                     \textcolor{blue}{\tcc{conflicts mean incomplete treatment team/process, leakage/contradictions}}
                \textit{(i)} do $\mathbb{R}_{IC} \gets (\mathbb{R}_{IC} - \mathbb{ID}_{IC})$\;
                     \textcolor{blue}{\tcc{delete altered, terminated, and expired informed consent from repository}}
                \textit{(ii)} do $\mathbb{AR}_{IC} \gets (\mathbb{AR}_{IC} + \mathbb{ID}_{IC})$\;
                     \textcolor{blue}{\tcc{add altered, terminated, and expired  informed consent to archive}}
                    
                \eIf{$\mathbb{R}_{IC} \xleftarrow{} (\mathbb{R}_{IC} - \mathbb{ID}_{IC})$ \&\& $\mathbb{AR}_{IC} \xleftarrow{} (\mathbb{AR}_{IC}+ \mathbb{ID}_{IC})$ == True}{
                    
                    \KwRet \textit{success: $\mathbb{ID}_{IC}$ removed from $\mathbb{R}_{IC}$ and added to $\mathbb{AR}_{IC}$}\;
                            \textcolor{blue}{\tcc{$\mathbb{ID}_{IC}$ cannot be executed for authorization}}
                    }{
                         \KwRet \textit{error: $\mathbb{IC}_{New}$ is not added to $\mathbb{AR}_{IC}$}\;                
                    }
                           
             }{
            \KwRet \textit{error: $\mathbb{ID}_{IC}:$ cannot be removed}\;
                    \textcolor{blue}{\tcc{avoid leakage/contradictions}}
        } 
\end{algorithm}

\section{Consent Provenance Services} \label{sec:consent-services} \vspace{-0.8em}

The consent service offers patients concise, clear, consistent, unmodified, real-time, and informative insights into both given and executed consent. Patients need to know to whom they have given consent, for what specific purposes, involving which resources, and under which conditions. Furthermore, patients should clearly understand how their consent is executed, including details such as who performs which operations and at what time. To ensure transparency and accountability, the service provides various assurances regarding the consent given and acted upon. This section explores the consent services tailored for patients within the proposed system, focusing on services oriented about users, resources, operations, and conditions \cite{albalwy2021blockchain}.

\begin{figure*}[hbt!]
    \centering
    \includegraphics[width=\textwidth]{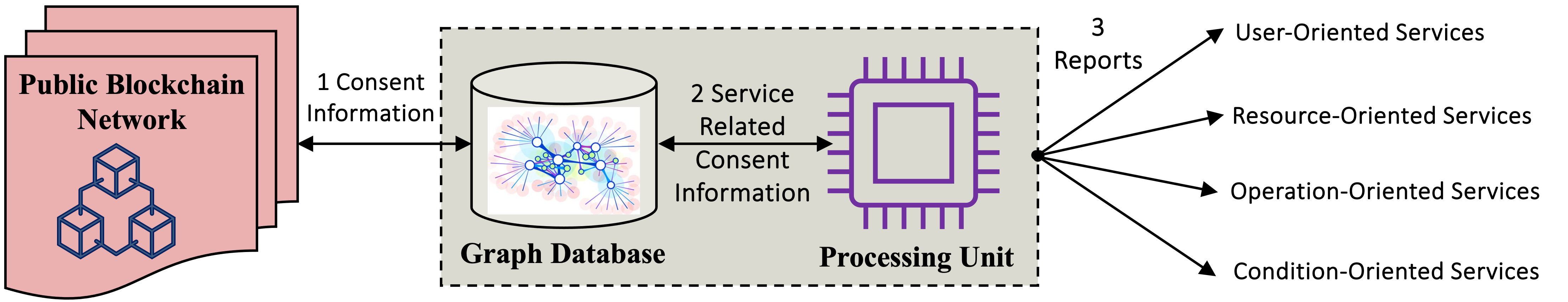}
    \vspace{-2em}
    \caption{Proposed Graph Database Based Consent Service Providing Mechanism.} \label{fig:graph-database-based-consent-services}
\end{figure*}

\textbf{Consent Services Mechanism: } A consent provenance service mechanism is proposed based on graph databases, as depicted in Fig. \ref{fig:graph-database-based-consent-services}. It initiates by collecting comprehensive informed consent information, including consent, related events, execution times, and more, from the public blockchain network (Step 1). This data is then stored in a graph database for further processing (Step 2). The processing unit retrieves consent-related information upon service requests and generates detailed reports (Step 3). These reports provision to various service orientations: (i) user-oriented, (ii) resource-oriented, (iii) operation-oriented, and (iv) condition-oriented services. A trusted and secured API, or Oracle, facilitates data acquisition from the blockchain network and subsequent storage in the graph database. This setup ensures ongoing monitoring of patient-related smart contract activities on the blockchain, capturing information for processing. Utilizing a graph database facilitates consent services by effectively handling complex relationships among patients, consents, and healthcare events, enabling simplified data retrieval and insightful visualization of consent patterns \cite{robinson2015graph}.

\begin{figure}[htbp]
  \centering
  \begin{minipage}[b]{0.48\textwidth}
    \includegraphics[width=\textwidth]{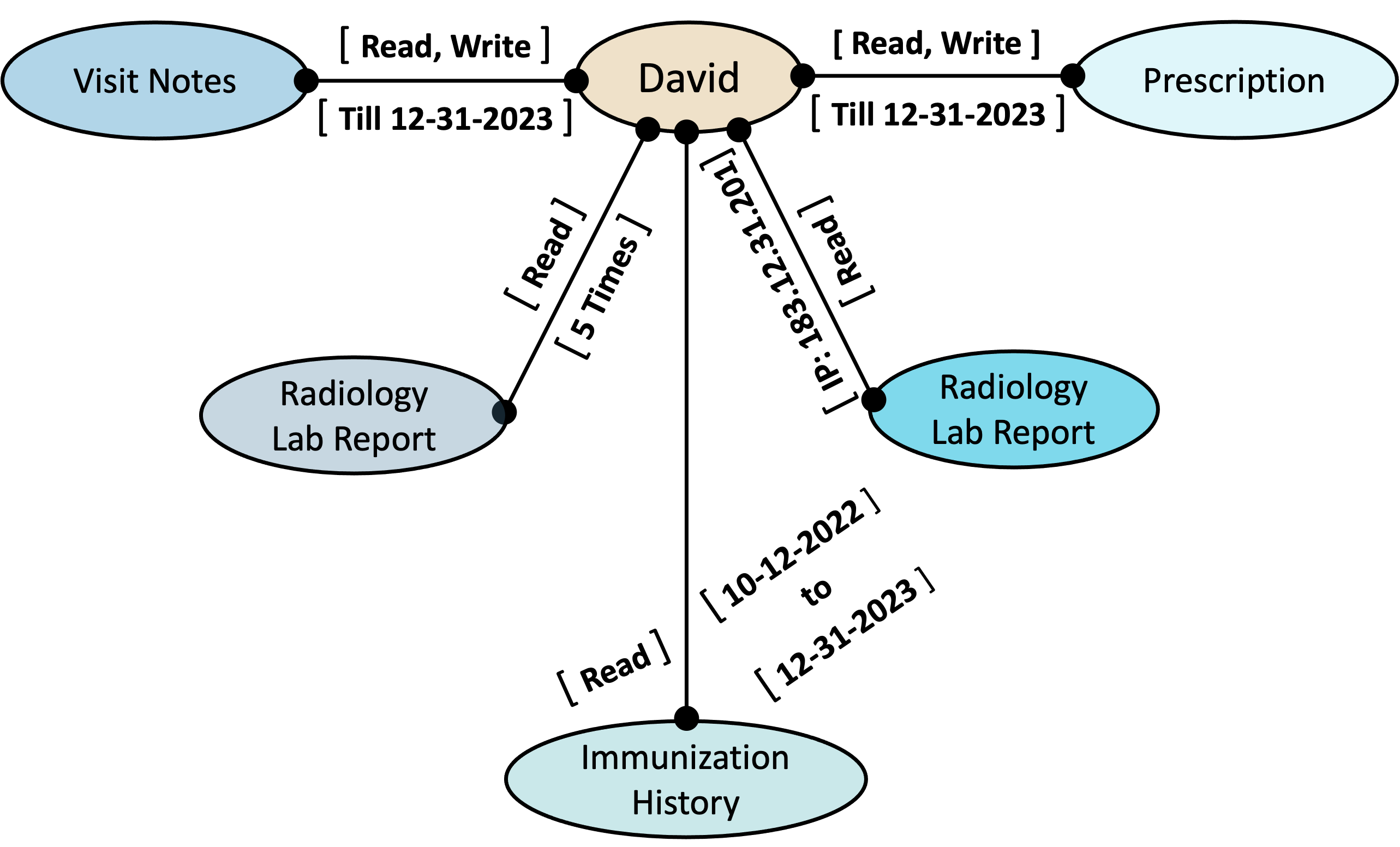}
    \vspace{-2em}
    \caption{User-Oriented Given Consents \cite{al2023informed}.} \label{fig:user-oriented-given-consents}
  \end{minipage}
  \hfill
  \begin{minipage}[b]{0.48\textwidth}
    \includegraphics[width=\textwidth]{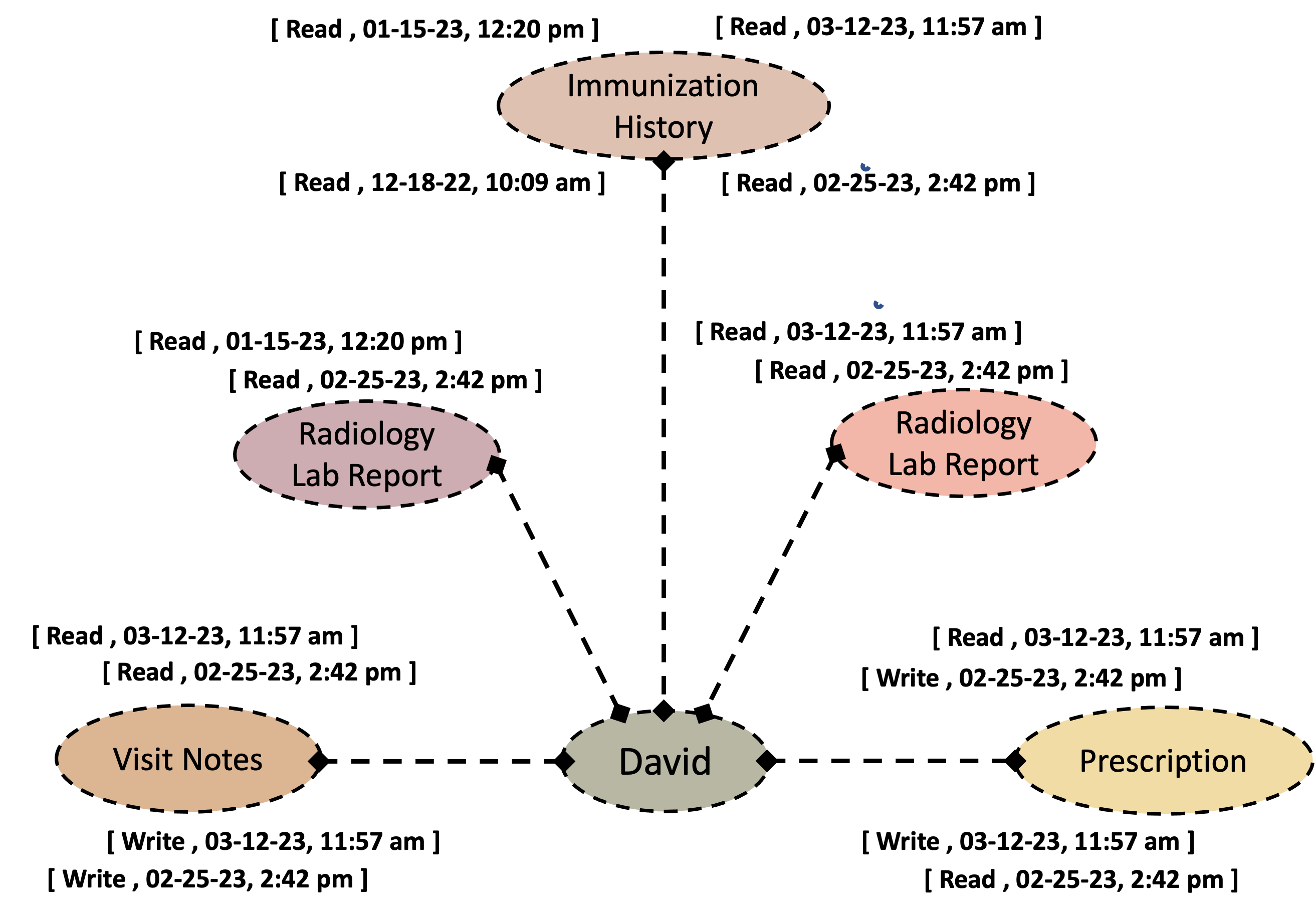}
    \vspace{-2em}
    \caption{User-Oriented Executed Consents \cite{al2023informed}.} \label{fig:user-oriented-executed-consents}
  \end{minipage}

    \begin{minipage}[b]{0.48\textwidth}
    \includegraphics[width=\textwidth]{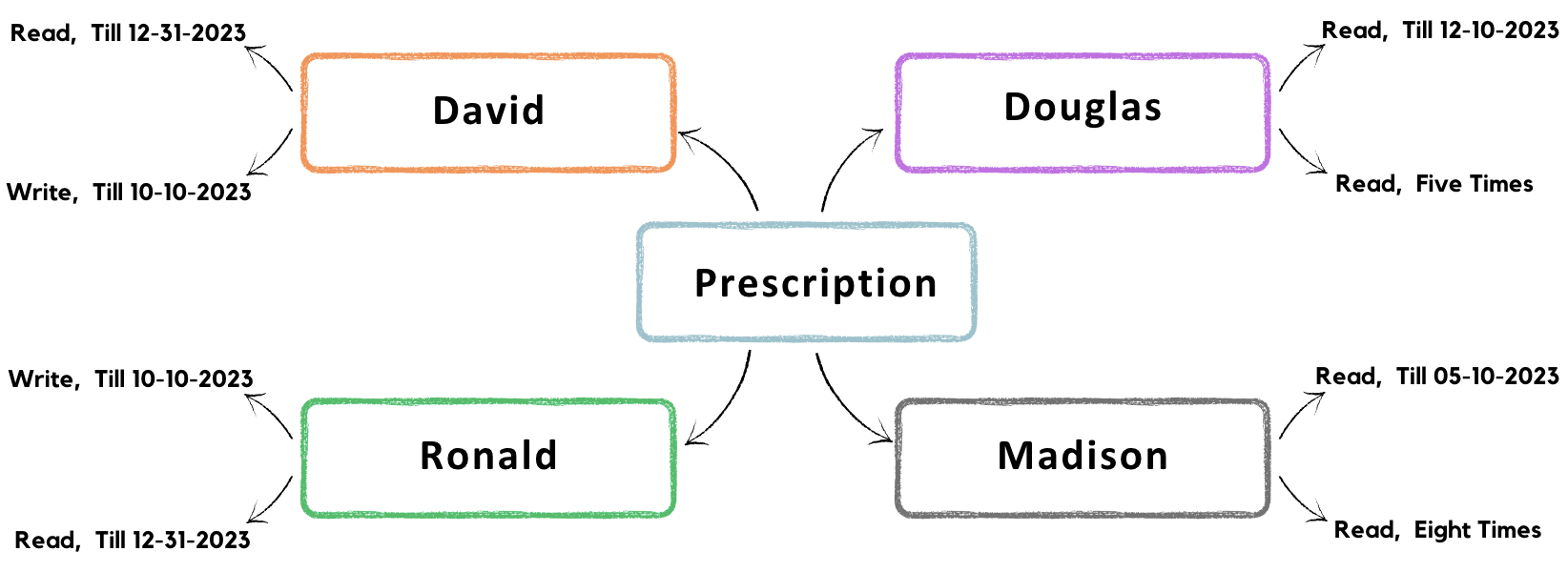}
    \vspace{-2em}
    \caption{Object-Oriented Given Consents \cite{al2023informed}.} \label{fig:object-oriented-given-consents}
  \end{minipage}
  \hfill
  \begin{minipage}[b]{0.48\textwidth}
    \includegraphics[width=\textwidth]{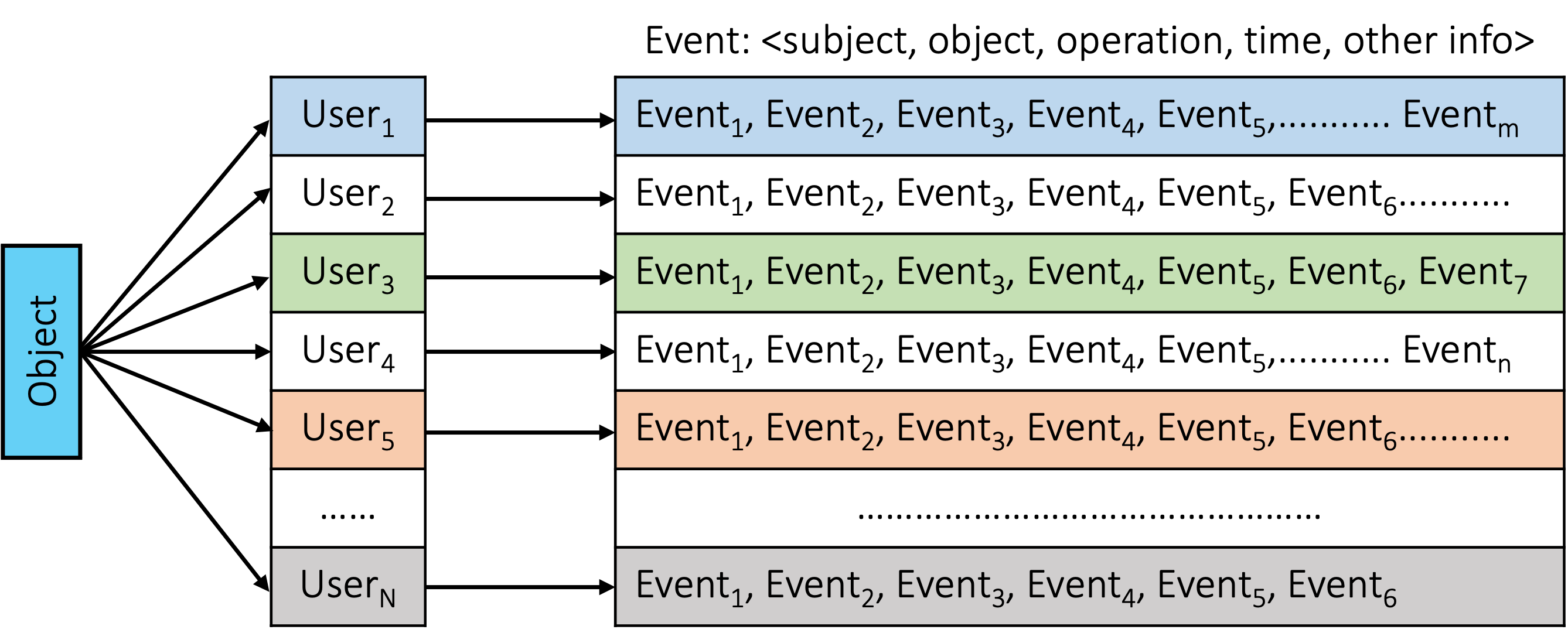}
    \vspace{-2em}
    \caption{Object-Oriented Executed Consents \cite{al2023informed}.} \label{fig:object-oriented-executed-consents}
  \end{minipage} 
\end{figure}

\textbf{User-Oriented Services: }  Patients can track specific users' consents, viewing a list of resources they've authorized for user operations, along with applicable conditions like access frequency and duration. This enables patients to audit any actions taken with their resources, ensuring transparency. Fig. \ref{fig:user-oriented-given-consents} illustrates the consents granted by patient $Jordan$ to doctor $David$, covering resources: \textit{Visit Notes, Prescription, Radiology Lab Report, Pathology Lab Report}, and \textit{Immunization History} detailed with operations and conditions. Furthermore, Fig. \ref{fig:user-oriented-executed-consents} displays the executed consents with operations, frequency, and access timing.

\textbf{Resource-Oriented Services: } Patients may require information on consents granted and executed for specific resources. The object-oriented consent service outlines all permissions, detailing who is authorized for which operations and under what conditions. Fig. \ref{fig:object-oriented-given-consents} presents a sample of such permissions, including the operations and conditions associated with each user and resource. Similarly, Fig. \ref{fig:object-oriented-executed-consents} illustrates the actual usage of these permissions, showing various events with details on who performed what action, when, and other information.

\textbf{Operation-Oriented Services: } This service provides detailed reports on both granted and executed consents for operations such as (i) read, (ii) write, and (iii) update. While reading operations do not affect data integrity, they may pose a risk to confidentiality in cases of unauthorized access. Conversely, write and update operations have the potential to alter data integrity. Ensuring that only authorized users and actions can modify data integrity is important.

\textbf{Conditions-Oriented Services: } Several conditions must be met for consent to be enforced for PHI access authorizations. Patients must be assured that these conditions are thoroughly verified. In this service mode, detailed information on both granted and executed consents is provided, focusing specifically on the associated conditions. This allows comprehensive visibility into how all included conditions are addressed and evaluated for making authorizations.

\section{Experimental Evaluation} \label{sec:experimental-evaluation} \vspace{-0.8em}

The Ethereum Virtual Machine (EVM) based blockchains are chosen for the proposed approach experiments. It offers a Turing-complete smart contract language, Solidity, which enables the implementation of our model's logic. We developed smart contracts for storing and retrieving informed consent, testing them on test networks: \textit{Arbitrum}, \textit{Polygon}, and \textit{Optimism} to ensure reliability before deployment. Since smart contracts, once deployed, are immutable and errors can incur financial and reputational costs, rigorous testing on these networks is crucial. Ethereum's Remote Procedure Call (RPC) API services are employed for deploying smart contracts on these test networks \cite{kim2023etherdiffer}. Utilizing public RPC, a toolkit for blockchain application development, eliminates the need to maintain a blockchain node for contract interaction, assuming minimal resource usage (CPU, HDD, bandwidth) on the local machine. Faucet \textit{ETH} and \textit{MATIC} serve as gas to authorize transactions using the Metamask wallet \cite{lee2023using}. 

\begin{figure}[tbp]
  \centering
  \begin{minipage}[b]{0.48\textwidth}
    \includegraphics[width=\textwidth]{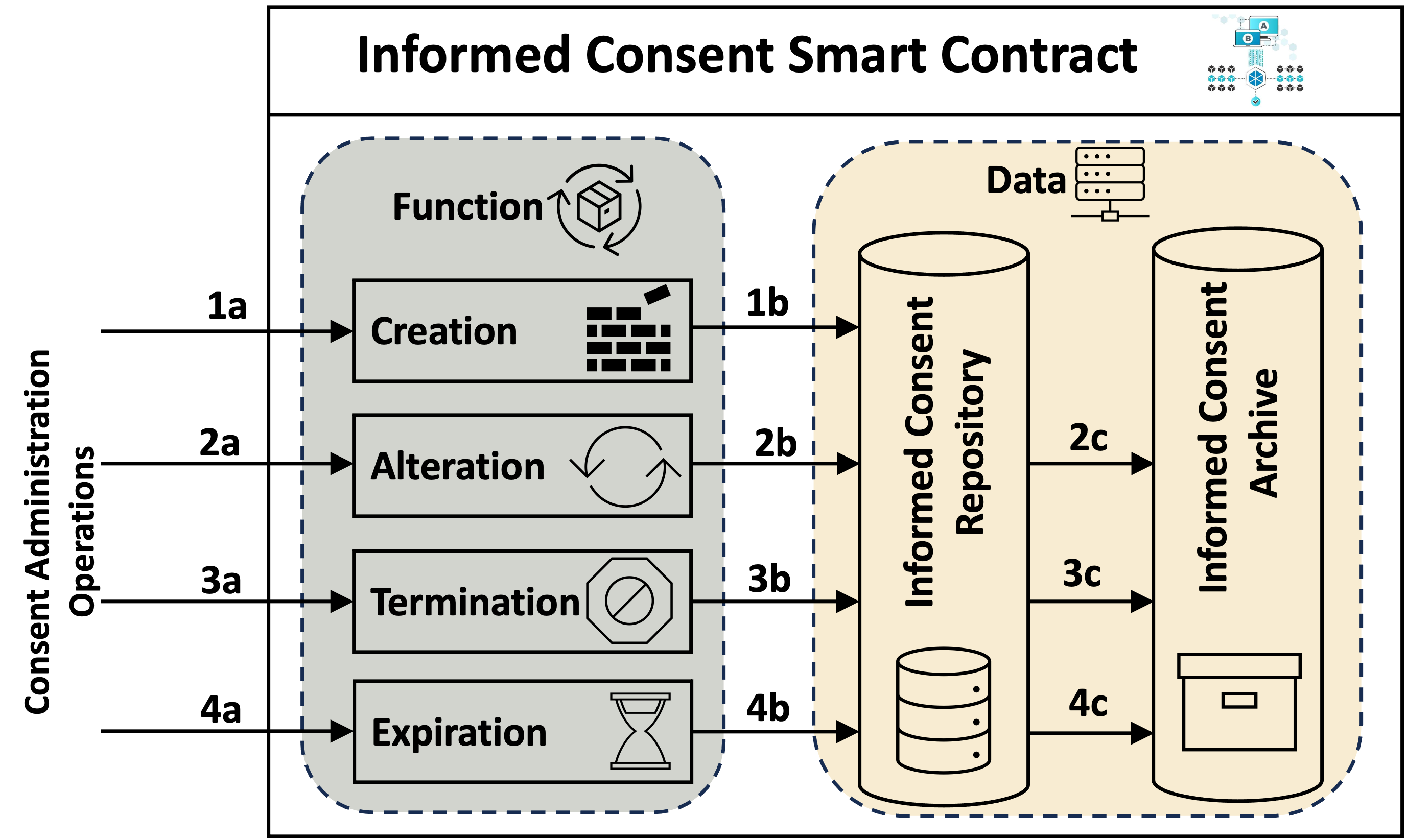}
    \vspace{-2em}
    \caption{Smart Contract Structure.} \label{fig:patient-smart-contract-structure}
  \end{minipage}
  
  \hfill
  
  \begin{minipage}[b]{0.48\textwidth}
    \includegraphics[scale=0.165]{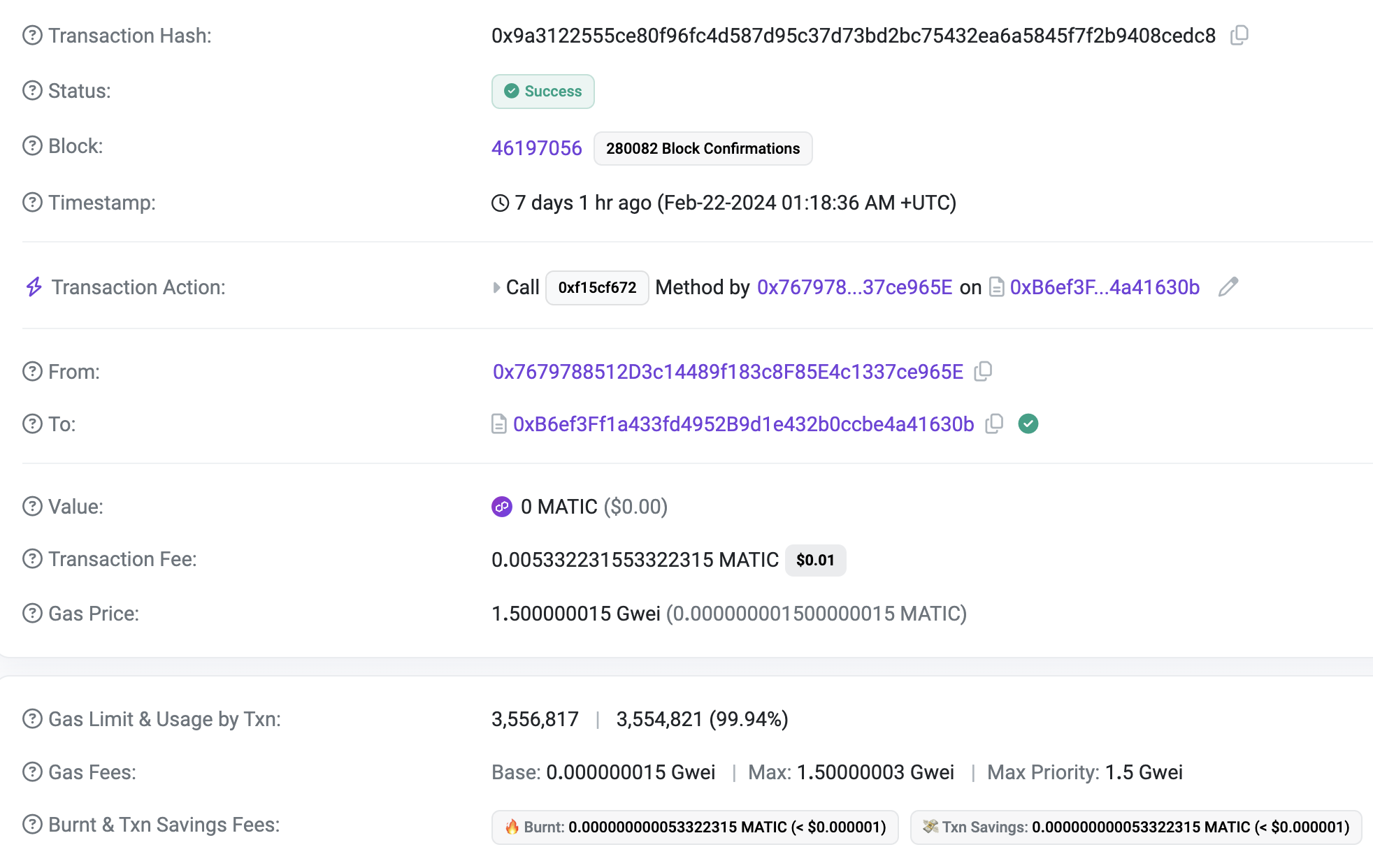}
    \vspace{-1em}
    \caption{Informed Consent Transaction.} \label{fig:informed-consent-transactions}
  \end{minipage}
\end{figure}

Fig. \ref{fig:patient-smart-contract-structure} shows the smart contract structure that acts as a container. Each smart contract has functions and data as storage. Functions perform particular operations: consent creation, alteration, termination, and expiration. The contract also stores consent as data. There are two scopes of storage: an informed consent repository that contains active consent that is executable for authorizations and an informed consent archive that contains historical consent from consent alteration, termination, and expiration operations. The archive provides a read-only repository, which means consent from here can not be executed. A transaction is depicted in Fig. \ref{fig:informed-consent-transactions} from Blockchain Explorer. The following discusses gas consumption and time requirements to assess the functionalities of the proposed approach for three test networks: \textit{Polygon}, \textit{Arbitrum}, and \textit{Optimism}.

\subsection{Smart Contract Deployment and Consents Storage Cost} \vspace{-.5em}

Gas is needed for any public Blockchain activity involving writing or changing data \cite{wood2014ethereum}. Some functions are sending ether (or any other ERC20 token), minting and sending NFTs, deploying smart contracts, changing the state of the blockchain, and so on. For this work, we only need to consider smart contract deployment and function calling costs to write data on the blockchain network. How much it costs to call a function depends on how many times it is called and how much data needs to be stored or changed on the blockchain network. 

\begin{figure}[htb]
    \centering
    \includegraphics[scale=0.22]{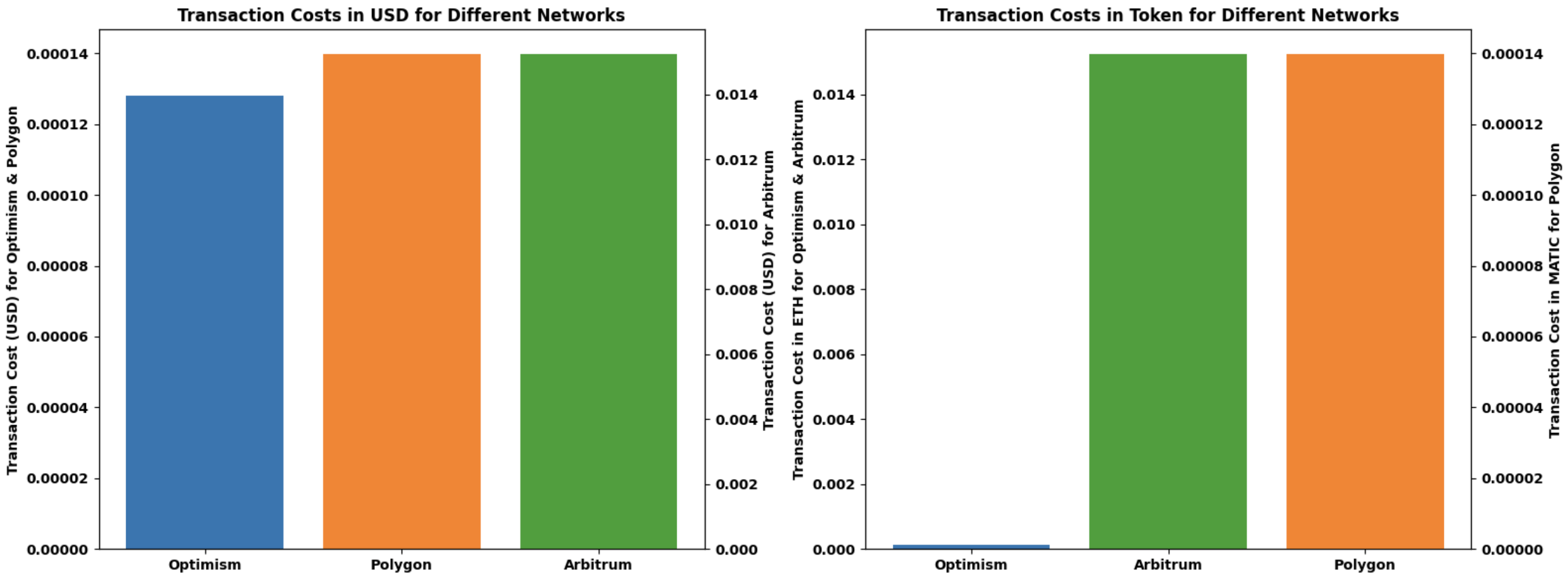}
    \vspace{-2em}
    \caption{PPA Integrity Writing Cost.}
    \label{fig:ppa-integrity-write-gas-cost}
\end{figure}

\textbf{PPA Integrity Storage Cost: } Fig. \ref{fig:ppa-integrity-write-gas-cost} displays two side-by-side bar graphs comparing transaction costs for \textit{PPAIngerityContract} smart contract deployed in \textit{Optimism, Polygon}, and \textit{Arbitrum} test networks. The size of a PPA hash as integrity is 32 bytes. The left graph shows costs in USD, while the right graph presents costs in native tokens (\textit{ETH} for Optimism and Arbitrum, \textit{MATIC} for Polygon). A clear trend is evident: Arbitrum's transaction costs are substantially higher than those of Optimism and Polygon, with its bars reaching the upper limits of the graphs. This suggests that users may face significantly higher fees on the Arbitrum network, which could influence their choice of platform for transactions or smart contract interactions.

\begin{figure}[htb]
    \centering
    \includegraphics[scale=0.18]{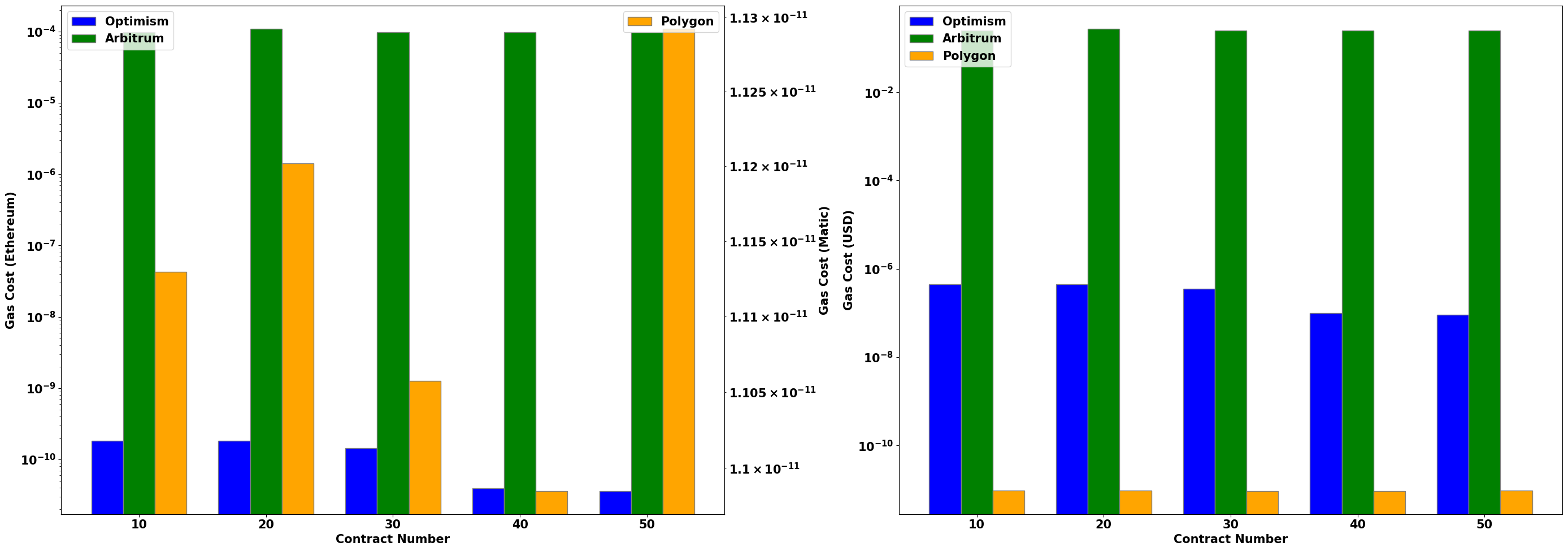}
    \vspace{-2em}
    \caption{Patient Informed Consent Smart Contract Deployment Cost}
    \label{fig:contract-deployment-cost}
\end{figure}

\textbf{Patient Smart Contract Deployment Cost: } Fig.\ref{fig:contract-deployment-cost} displays two bar charts comparing the gas costs for deploying identical smart contracts across three test networks. The left chart uses a logarithmic scale on the y-axis to showcase the huge difference in deployment costs, particularly highlighting how Arbitrum incurs higher costs than Optimism. The variation in deployment costs is attributed to the cost of native tokens and network congestion levels at the deployment time. These factors contribute to the differing costs despite identical contract codes, as illustrated in the figure.

\begin{figure}[tb]
    \centering
    \includegraphics[scale=0.3]{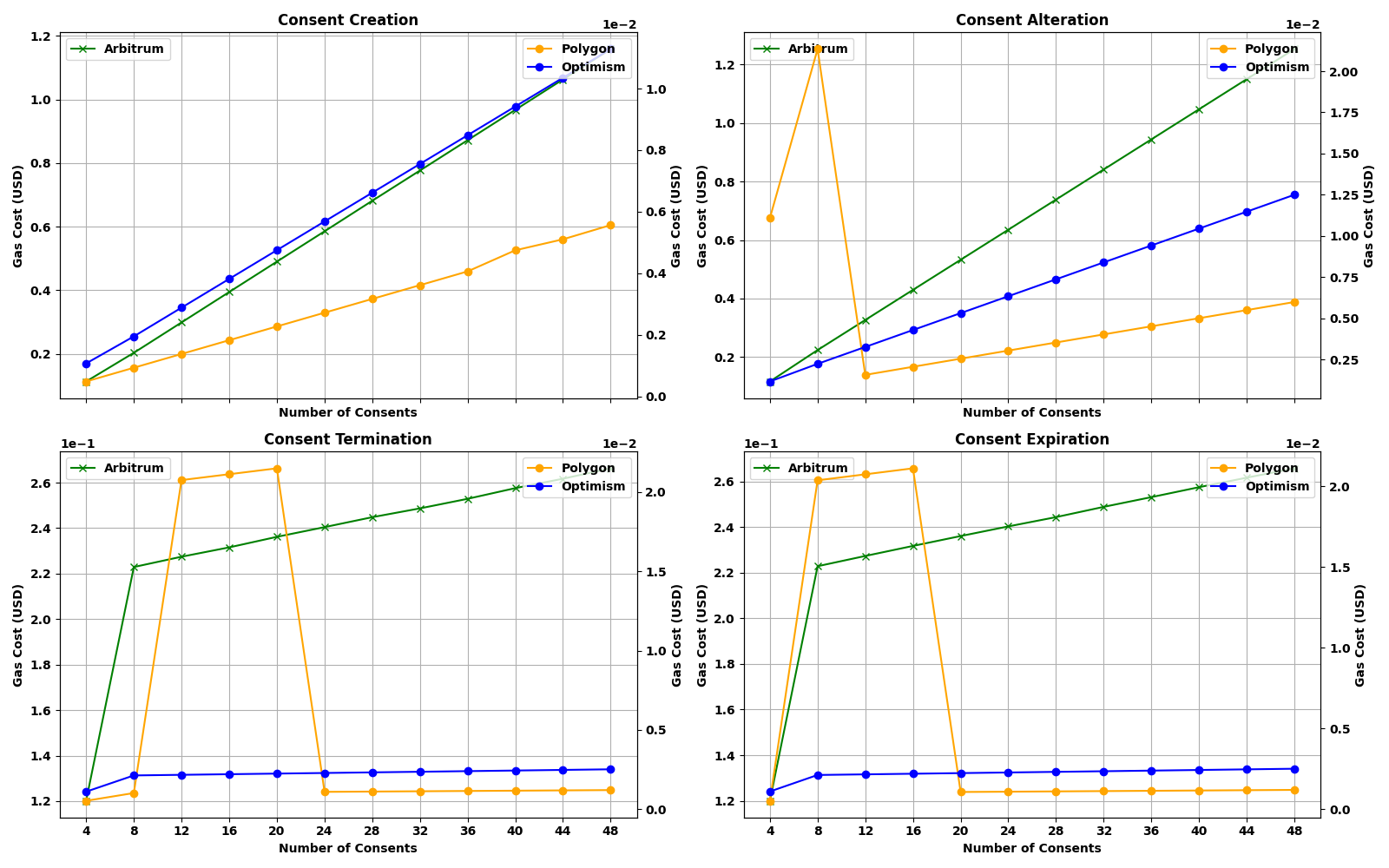}
    \vspace{-3em}
    \caption{Consent Creation, Alteration, Termination, and Expiration Cost.}
    \label{fig:operations-cost}
\end{figure}

\textbf{Consent Creation, Alteration, Termination, and Expiration Cost: }  The creation operation involves writing new consents to the active consent repository. The other operations—alteration, termination, and expiration—need the transfer of active consents to the read-only archive, effectively changing their status from active to historical. Fig.\ref{fig:operations-cost} illustrates the variation in transaction fees for different operations as the number of consents increases on three test networks. The volatility observed in these graphs can be attributed to network congestion, yet the price differences remain minimal. There is a noticeable gradual cost increase correlating with the rise in consent. Using scientific notation on the graph’s scales facilitates uniform axis labeling. It provides a coherent point of comparison for vastly different values, with the power denoted at the top for reference.

\subsection{Informed Consents Writing and Reading Time Requirements} \vspace{-.5em}

All the read calls of smart contracts are gas-free. Smart contract deployment and execution stages are the basis of the time cost associated with on-chain activities. Any blockchain-based applications require two kinds of time requirements: \textit{(i) block data writing} and \textit{(ii) block data reading}. 

\textbf{Writing Time:}  Writing time includes smart contract deployment and adding data. A new block is added to the Ethereum main network every 12 seconds on average, ideal for the proposed purposes \cite{pierro2019influence}. So long as there is sufficient space in new blocks, a new transaction would take, on average, no more than 12 seconds. If block congestion occurs, the time it takes for a transaction to be included in a block might increase. However, users can influence this by paying more gas for faster block confirmation. Given that users may artificially extend the confirmation time of their transactions. Table \ref{table:consent-writing-time} shows the writing time for various consent numbers for test networks. Table \ref{table:consent-administration-writing-time} depicts the same test networks' writing time consent administration operations: alteration, termination, and expiration. These operations require moving consents from the active repository to a read-only archive. For both tables, it is noticeable that \textit{Arbitrum} requires less time than the other two networks. This is because of the sequencer design and network congestion management \cite{kalodner2018arbitrum}. The same smart contracts and consents are used for all test networks.

\textbf{Reading Time:} The reading time indicates the required time to get data from the block of the blockchain ledger. All the read calls of smart contracts are gas-free. Table \ref{table:consent-reading-time} depicts the reading time for various consent numbers for test networks. For the same test networks, the reading time for consent administration operations is tabulated in Table \ref{table:consent-administration-reading-time}. Maintaining a node locally can reduce reading time from the network. Where block data can be accessed in real-time. The system continuously synchronizes with the blockchain network to update the ledger data. The hospital authorities can maintain local nodes for faster authorization decisions.

\begin{table}[htb]
  \centering
  \begin{minipage}[b]{0.48\textwidth}
   \caption{Consent Writing Time.} \label{table:consent-writing-time} \vspace{-.5em}
\scriptsize
\begin{tabular}{|>{\centering\arraybackslash}p{0.21\linewidth}|>{\centering\arraybackslash}p{0.22\linewidth} |>{\centering\arraybackslash}p{0.24\linewidth}| >{\centering\arraybackslash}p{0.24\linewidth}|}
\hline
   \textbf{Consents} &  \textbf{Polygon} & \textbf{Optimism} & \textbf{Arbitrum}  \\
\hline
     4 & 5.708 Sec & 2.744 Sec  & 1.885 Sec\\ 
 \hline 
    8 & 6.689 Sec & 2.702 Sec &  1.633 Sec\\
\hline
    12 & 7.092 Sec & 3.562 Sec & 6.825 Sec \\
\hline
    16 & 5.418 Sec &  8.464 Sec &  3.181 Sec\\
\hline
     20 & 7.448 Sec & 7.363 Sec  & 1.586 Sec \\ 
 \hline 
    24 & 5.457 Sec & 8.375 Sec &  5.778 Sec\\
\hline
    28 & 6.772 Sec & 7.805 Sec  & 1.73 Sec\\
\hline
    32 & 5.972 Sec & 7.943 Sec  &  3.39 Sec\\
\hline
    36 & 5.542 Sec & 7.736 Sec &  1.834 Sec\\
\hline
    40 & 6.128 Sec & 7.573 Sec  &  2.119 Sec\\
\hline
    44 & 7.536 Sec & 7.39 Sec &  7.536 Sec \\
\hline
    48 & 5.521 Sec & 7.698 Sec &  3.394 Sec\\
\hline
\end{tabular}
  
  \end{minipage}
  \hfill
  \begin{minipage}[b]{0.48\textwidth}
    \caption{Consent Reading Time.} \label{table:consent-reading-time} \vspace{-.5em}
    \scriptsize

\begin{tabular}{|>{\centering\arraybackslash}p{0.21\linewidth}|>{\centering\arraybackslash}p{0.22\linewidth} |>{\centering\arraybackslash}p{0.24\linewidth}| >{\centering\arraybackslash}p{0.24\linewidth}|}
\hline
 \textbf{Consents} &  \textbf{Polygon} & \textbf{Optimism} & \textbf{Arbitrum}  \\
\hline
     4 & 0.466 Sec & 0.228 Sec  &  0.427 Sec \\ 
 \hline 
    8 & 0.472 Sec & 0.52 Sec &  0.289 Sec \\
\hline
    12 & 0.497 Sec & 0.208 Sec &  0.727 Sec\\
\hline
    16 & 0.591 Sec & 0.201 Sec  & 0.975 Sec \\
\hline
     20 & 0.504 Sec & 0.223 Sec  & 0.33 Sec \\ 
 \hline 
    24 & 0.923 Sec & 0.221 Sec &  0.304 Sec \\
\hline
    28 & 0.6 Sec & 0.235 Sec &  0.305 Sec\\
\hline
    32 & 0.909 Sec & 0.245 Sec  & 0.32 Sec\\
\hline
    36 & 0.526 Sec & 0.229 Sec &  0.719 Sec  \\
\hline
    40 & 0.812 Sec & 0.257 Sec   & 0.363 Sec\\
\hline
    44 & 0.742 Sec &  0.457 Sec  &0.247 Sec \\
\hline
    48 & 0.631 Sec & 0.557 Sec  & 0.266 Sec \\
\hline
\end{tabular}
  \end{minipage}
\end{table}

\begin{table}[htb]
  \centering
\begin{minipage}{\linewidth}
\centering
\caption{Writing Time for Consent Administration Operation in Seconds.} \label{table:consent-administration-writing-time}
\scriptsize
\vspace{-.5em}
\begin{tabular}{|c|c|c|c|c|c|c|c|c|c|}
\hline
 \multirow{2}{*}{Consents} & \multicolumn{3}{c|}{Alteration Operation} & \multicolumn{3}{c|}{Termination Operation} & \multicolumn{3}{c|}{Expiration Operation} \\
 \cline{2-10}
 & \multicolumn{1}{c|}{Polygon} & \multicolumn{1}{c|}{Optimism} & \multicolumn{1}{c|}{Arbitrum} & \multicolumn{1}{c|}{Polygon} & \multicolumn{1}{c|}{Optimism} & \multicolumn{1}{c|}{Arbitrum} & \multicolumn{1}{c|}{Polygon} & \multicolumn{1}{c|}{Optimism} & \multicolumn{1}{c|}{Arbitrum} \\
\hline
4 & 6.610 & 7.109 & 2.597 & 6.705 & 6.838 & 2.574 & 6.665 & 7.050 & 2.597 \\
\hline
8 & 6.670 & 6.967 & 2.465 & 6.769 & 6.874 & 2.284 & 6.592 & 6.916 & 2.210 \\
\hline
12 & 6.671 & 6.962 & 2.484 & 6.860 & 6.979 & 2.787 & 6.622 & 6.842 & 2.307 \\
\hline
16 & 7.024 & 2.871 & 2.240 & 6.667 & 6.946 & 2.349 & 6.729 & 6.839 & 2.485 \\
\hline
20 & 6.926 & 6.974 & 2.327 & 6.826 & 7.126 & 2.552 & 6.697 & 6.928 & 2.127 \\
\hline
24 & 6.739 & 7.066 & 2.562 & 6.732 & 7.053 & 2.849 & 6.850 & 6.875 & 2.784 \\
\hline
28 & 6.839 & 7.022 & 2.486 & 10.774 & 2.848 & 2.304 & 7.232 & 2.876 & 2.418 \\
\hline
32 & 6.797 & 7.128 & 2.809 & 6.853 & 6.862 & 2.299 & 6.581 & 7.067 & 2.324 \\
\hline
36 & 6.862 & 7.176 & 3.127 & 6.714 & 6.839 & 2.361 & 6.613 & 7.266 & 2.687 \\
\hline
40 & 6.942 & 7.630 & 2.533 & 6.683 & 6.958 & 2.602 & 6.658 & 7.084 & 2.414 \\
\hline
44 & 7.000 & 7.011 & 2.886 & 6.680 & 6.884 & 2.166 & 10.655 & 6.760 & 2.163 \\
\hline
48 & 6.948 & 7.195 & 2.597 & 6.891 & 7.036 & 2.409 & 6.818 & 2.217 & 2.548 \\
\hline
\end{tabular}

\vspace{1em}

\caption{Reading Time for Consent Administration Operation in Seconds.}  \label{table:consent-administration-reading-time}
\scriptsize
\vspace{-.5em}
\begin{tabular}{|c|c|c|c|c|c|c|c|c|c|}
\hline
\multirow{2}{*}{Consents} & \multicolumn{3}{c|}{Alteration Operation} & \multicolumn{3}{c|}{Termination Operation} & \multicolumn{3}{c|}{Expiration Operation} \\
\cline{2-10}
& \multicolumn{1}{c|}{Polygon} & \multicolumn{1}{c|}{Optimism} & \multicolumn{1}{c|}{Arbitrum} & \multicolumn{1}{c|}{Polygon} & \multicolumn{1}{c|}{Optimism} & \multicolumn{1}{c|}{Arbitrum} & \multicolumn{1}{c|}{Polygon} & \multicolumn{1}{c|}{Optimism} & \multicolumn{1}{c|}{Arbitrum} \\
\hline
4 & 0.417 & 0.466 & 0.482 & 0.381 & 0.419 & 0.401 & 0.410 & 0.427 & 0.395 \\
\hline
8 & 0.426 & 0.419 & 0.472 & 0.411 & 0.411 & 0.427 & 0.398 & 0.401 & 0.405 \\
\hline
12 & 0.424 & 0.418 & 0.480 & 0.403 & 0.438 & 0.389 & 0.405 & 0.408 & 0.429 \\
\hline
16 & 0.602 & 0.547 & 0.462 & 0.560 & 0.485 & 0.399 & 0.422 & 0.420 & 0.399 \\
\hline
20 & 0.479 & 0.528 & 0.503 & 0.551 & 0.538 & 0.482 & 0.463 & 0.624 & 0.461 \\
\hline
24 & 0.508 & 0.714 & 0.465 & 0.453 & 0.482 & 0.537 & 0.541 & 0.574 & 0.515 \\
\hline
28 & 0.564 & 0.639 & 0.566 & 0.476 & 0.478 & 0.481 & 0.515 & 0.672 & 0.467 \\
\hline
32 & 0.632 & 0.563 & 0.629 & 0.514 & 0.504 & 0.449 & 0.493 & 0.513 & 0.501 \\
\hline
36 & 0.685 & 0.657 & 0.632 & 0.487 & 0.495 & 0.552 & 0.484 & 0.555 & 0.590 \\
\hline
40 & 0.832 & 0.859 & 0.674 & 0.499 & 0.513 & 0.811 & 0.476 & 0.632 & 0.601 \\
\hline
44 & 0.890 & 0.753 & 0.642 & 0.495 & 0.504 & 0.473 & 0.528 & 0.494 & 0.556 \\
\hline
48 & 1.197 & 0.838 & 0.639 & 0.494 & 0.501 & 0.547 & 0.552 & 0.515 & 0.559 \\
\hline
\end{tabular}
\end{minipage}
\end{table}

\section{Additional Factors Consideration} \label{sec:additional-factors} \vspace{-0.9em}
Healthcare providers must consider providing services to patients and cost coverage scopes in addition to other points mentioned in this proposed approach. They are important to provide treatment and healthcare services to patients.

\textbf{Patient Service Delivery: }  Patients engage with the healthcare system through user-friendly interfaces like GUIs, applications, or simplified platforms to access services. Wallet applications such as Coinbase and MetaMask facilitate interaction with the blockchain by signing transactions and managing digital currencies or tokens, securely storing private keys and credentials for users \cite{he2020security}. The healthcare providers must accommodate diverse user needs, including those requiring tailored software interfaces or apps for ease of use, such as older adults, individuals with physical disabilities, minors, and others with limited IT knowledge or specific requirements \cite{mtshali2019smart}. This research concentrates on developing methods to capture, store, and enforce patient informed consent for medical diagnosis and treatment. It's envisioned that healthcare providers will address the unique needs of their patients, ensuring that patients are not versed in complex underlying technologies like blockchain, smart contracts, or distributed systems. We assume that patient devices and applications are safeguarded against unauthorized access and that communication between these devices/apps and the blockchain network remains secure.

\textbf{System Operational Costs:} Some blockchain-based frameworks need transaction fees, like Gas in Ethereum. The gas consumption or transaction fee is considered for research and technical aspects, not from the patients' or users' perspective. Healthcare providers can spend on infrastructure expenses such as blockchain network nodes, apps for mobile devices to interact with hospitals, blockchain systems, and others. There are direct costs regarding storing informed consent on public blockchain networks like Ethereum. The patients, insurance companies, and others can cover these costs, like doctors' fees, medications, pathology lab tests, radiology lab tests, and other direct/indirect costs related to treatment. In blockchain networks, state change operations require spending money, while reading from the network does not need monetary expenditure. Once informed consents are deployed to the blockchain networks, relevant users can access them without spending charges.

\section{Literature Review} \label{sec:literature-review} \vspace{-0.8em}

Numerous proposals have been made for integrating blockchain technology into healthcare and e-health systems. Existing research primarily focuses on using blockchain to safeguard medical records and facilitate the storage and sharing of medical data, analytics, and systems for managing informed consent in clinical or research settings. There has been some research dedicated to informed consent in diagnosis and treatment as well \cite{ploug2012pharmaceutical}. To the best of our knowledge, our study is the inaugural effort to leverage blockchain and smart contracts specifically to manage and enforce informed consent in clinical diagnosis and treatment.

Azaria et al. \cite{azaria2016medrec} introduced \textit{MedRec}, a healthcare data management system built on blockchain technology to improve access and permissions for electronic medical records. This system tackles critical challenges such as fragmented access to medical data, a lack of system interoperability, limited patient control over their information, and the need for enhanced data quality and quantity for research purposes. MedRec provides patients with a comprehensive, immutable record of their medical history, facilitating easy retrieval of information from healthcare providers and treatment facilities. By aggregating and encoding references to various types of medical data on a blockchain ledger, MedRec establishes a transparent and accessible historical trail for medical information. Cunningham et al. \cite{cunningham2022non} introduced the concept of using \textit{Non-Fungible Tokens (NFTs)} to document and transfer patient consent records for utilizing medical data. This approach allows individuals to record consent agreements, thereby authorizing \textit{Data Consumers} to access medical information from \textit{Data Providers} according to the permissions granted by the data's subjects. Nevertheless, the application of NFTs in tracking data provenance in compliance with regulatory standards like \textit{HIPAA/GDPR} is still under exploration. 

Albalwy et al. \cite{albalwy2021blockchain} introduced a blockchain-based consent management system, ConsentChain, designed to enhance the sharing of clinical genomic data. Utilizing the Ethereum blockchain, it employs smart contracts to represent the roles and permissions of patients (who grant or revoke access to their data), data creators (who gather and maintain patient information), and data requesters (who seek access to this information). While this initiative focuses on facilitating the exchange of genomic data among clinicians, researchers, and bioinformaticians, it's noted that clinical treatment presents distinct challenges for consent management compared to genomic data sharing. The treatment process involves various user actions, such as reading, writing, and modifying data, with access rights tailored to each user's role. To address the nuanced requirements of managing permissions for various clinical treatment stakeholders— from treatment team members to insurance agents and pharmacists—we propose a consent management framework specifically designed to handle these complex scenarios.

Tith et al. \cite{tith2020patient} designed an e-consent management system leveraging the Hyperledger Fabric blockchain alongside a purpose-based access control framework. This system records all patient data, consents, and metadata concerning data access on the blockchain, making it accessible to participating organizations. A specific chaincode executes the business logic for handling patient consent, allowing patients to initiate, modify, or revoke their consent directly on the blockchain. While this model is also suitable for data donations to biobank research, Hyperledger—a permissioned blockchain—limits participation to specific organizations, potentially reducing transparency to the broader public. In contrast, our approach utilizes public blockchain networks like Ethereum, inviting participants with stakes to join and uphold the ledger, thus offering immutable information even to untrusted parties. The public consensus mechanism ensures greater transparency compared to permissioned networks. Furthermore, Ethereum's smart contracts, which are widely utilized and continuously improved upon, present a more established and versatile platform for development.

Yue et al. \cite{yue2016healthcare} introduced a blockchain-powered application, Healthcare Data Gateway, enabling patients to possess, manage, and distribute their healthcare information securely. The system facilitates the secure processing of healthcare data by untrusted parties through the use of secure multi-party computation techniques, ensuring the protection of patient privacy. Xia et al. \cite{xia2017bbds} developed a framework for sharing data through blockchain technology, tackling the issue of access control for sensitive information in cloud storage. This approach leverages blockchain's inherent immutability and autonomous features to ensure secure data access. Zyskind et al. \cite{zyskind2015decentralizing} introduced the use of blockchain technology for managing access control and securely storing data, with encryption used to protect the data stored on servers. Fan et al. \cite{fan2018medblock} proposed MedBlock, a blockchain-driven information management system to streamline access and retrieval of electronic medical records (EMR). It ensures user privacy through tailored access control and encryption mechanisms during data sharing.

\section{Conclusions and Future Directions} \label{sec:conclusion-future-directions} 
\vspace{-0.6em}
The acquisition of informed consent is vital to healthcare provision's ethical and pragmatic aspects. Patients must be thoroughly briefed, participate actively, and retain autonomy over their healthcare decisions. Healthcare providers must ensure patients know their treatment options to make well-informed decisions. This strengthens trust, empowers patients, and enhances the level of care. Informed consent is not merely a singular occurrence but a dynamic, enduring process that commences before any medical intervention and is maintained throughout the continuum of patient care, with patients retaining the right to amend or revoke their consent at any time.

Frameworks for managing consent through smart contracts are gaining recognition as effective means to address the secure and confidential handling of health data. These systems empower patients to manage their protected health information and to grant informed consent for healthcare professionals to access it. Blockchain technology enhances these frameworks, offering secure and streamlined consent management while providing the benefits of decentralization, transparency, and immutability. These features collectively boost audibility and accountability within healthcare data systems, making them a feasible solution for enhancing healthcare data accessing practices while upholding patient privacy.

Looking forward, our objective is to develop a comprehensive healthcare policy compliance framework that incorporates patient-informed consent and considers applicable policies and industry best practices for authorizing access to healthcare data by treatment team members. We plan to enhance the robustness of this framework by utilizing blockchain technology to capture and securely store audit trails as immutable records to provide provenance. This will enable the reliable detection of any unauthorized modifications, making tampering computationally infeasible. So that the activities can be recreated as they happened and captured. Additionally, we aim to refine our compliance verification process, using the provenance information to identify instances of both compliance and non-compliance. A blockchain consensus mechanism called Proof of Compliance (PoC) will be utilized to find compliance status. By implementing such a framework, healthcare organizations will be equipped to proactively identify and rectify non-compliance issues, thereby minimizing regulatory and legal penalties, safeguarding their reputation, and preventing financial and customer losses.

\section*{Acknowledgements} \vspace{-0.9em}
This work was partially supported by the U.S. National Science Foundation under Grant No. 1822118 and 2226232. Any opinions, findings, and conclusions or recommendations expressed in this material are those of the authors and do not necessarily reflect the views of the National Science Foundation, or other federal agencies.

\bibliographystyle{splncs04}
\bibliography{main}

\end{document}